\documentclass{aa}

\usepackage{graphicx, comment}
\usepackage{xcolor}
\usepackage{txfonts}
\usepackage[colorlinks=true]{hyperref}

\hypersetup{
     colorlinks   = true,
     citecolor    = blue
}

\usepackage{float}
\usepackage[flushleft]{threeparttable}
\usepackage{adjustbox}

\newcommand{\C}{\object{3C\,84}} 
\newcommand{\kgpr}{0.98\pm0.18} 
\newcommand{\kdcf}{1.21\pm0.22} 
\newcommand{\kav}{1.08\pm0.18} 
\newcommand{\znf}{22-645\,\textrm{R}_\textrm{s}} 
\newcommand{\znfpc}{2-20\times10^{-3}\,\textrm{pc}} 
\newcommand{\tauonemmgammat}{1.43\pm0.30} 
\newcommand{\tauonemmgammap}{1.56\pm0.27} 
\newcommand{\taupemmgammat}{1.58\pm0.64} 
\newcommand{\taupemmgammap}{1.57\pm0.49} 
\newcommand{\zonemm}{405-1622} 
\newcommand{\zpemm}{450-1802} 
\newcommand{\multpar}{3\times10^{12}-2\times10^{13}} 
\newcommand{\michel}{3-56} 
\newcommand{\Bno}{3.33-9.76} 
\newcommand{\Bnoround}{3-10}

\begin{document}

    \title{A multiband study and exploration of the radio wave - $\gamma$-ray connection in \C}

   \author{
   G.~F. Paraschos\inst{1}, 
   V. Mpisketzis\inst{11, 2}, 
   J.-Y. Kim\inst{10, 3, 1}, 
   G. Witzel\inst{1},
   T.~P. Krichbaum\inst{1},
   J.~A. Zensus\inst{1}, 
   M.~A. Gurwell\inst{4}, 
   A. L\"ahteenm\"aki\inst{5,6}, 
   M. Tornikoski\inst{5}, 
   S. Kiehlmann\inst{7,8}, 
   A.~C.~S. Readhead\inst{9}
          }
   \authorrunning{G.~F. Paraschos et al.}
   \institute{$^{1}$Max-Planck-Institut f\"ur Radioastronomie, Auf dem H\"ugel 69, Bonn, D-53121 Bonn, Germany\\ 
              $^{}$\ \email{gfparaschos@mpifr-bonn.mpg.de}\\
              $^{2}$Department of Physics, National and Kapodistrian University of Athens, Panepistimiopolis, GR 15783 Zografos, Greece\\
              $^{3}$Korea Astronomy and Space science Institute, 776 Daedeokdae-ro, Yuseong-gu, Daejeon, 30455, Republic of Korea\\
              $^{4}$Center for Astrophysics | Harvard \& Smithsonian, 60 Garden Street, Cambridge, MA 02138, USA\\
              $^{5}$Aalto University Mets\"ahovi Radio Observatory, Mets\"ahovintie 114, 02540 Kylm\"al\"a, Finland\\
              $^{6}$Aalto University Department of Electronics and Nanoengineering, P.O. BOX 15500, FI-00076 AALTO, Finland\\
              $^{7}$Institute of Astrophysics, Foundation for Research and Technology-Hellas, GR-71110 Heraklion, Greece\\
              $^{8}$Department of Physics, Univ. of Crete, GR-70013 Heraklion, Greece\\
              $^{9}$Owens Valley Radio Observatory, California Institute of Technology, Pasadena, CA 91125, USA\\
              $^{10}$Department of Astronomy and Atmospheric Sciences, Kyungpook National University, Daegu 702-701, Republic of Korea\\
              $^{11}$Institut f\"ur Theoretische Physik, Goethe Universit\"at Frankfurt, Max-von-Laue-Str.1, 60438 Frankfurt am Main, Germany
              }

   \date{Received -; accepted -}

  \abstract{
  Total intensity variability light curves offer a unique insight into the ongoing debate about the launching mechanism of jets.
  For this work, we utilise the availability of radio and $\gamma$-ray light curves over a few decades of the radio source \C\ (\object{NGC\,1275}). 
  We calculate the multiband time lags between the flares identified in the light curves via discrete cross-correlation and Gaussian process regression.
  We find that the jet particle and magnetic field energy densities are in equipartition ($k_\textrm{r} = \kav$).
  The jet apex is located $z_\textrm{91.5\,GHz}=\znf$ ($\znfpc$) upstream of the 3\,mm radio core; at that position, the magnetic field amplitude is $B_\textrm{core}^\textrm{91.5\,GHz}=\Bnoround$\,G.
  Our results are in good agreement with earlier studies, which utilised very-long-baseline interferometry.
  Furthermore, we investigate the temporal relation between the ejection of radio and $\gamma$-ray flares.
  Our results are in favour of the  $\gamma$-ray emission being associated with the radio emission.
  We are able to tentatively connect the ejection of features identified at 43 and 86\,GHz to prominent $\gamma$-ray flares.
  Finally, we compute the multiplicity parameter $\lambda$ and the Michel magnetisation $\sigma_\textrm{M}$ and find that they are consistent with a jet launched by the \cite{Blandford77} mechanism.
}

   \keywords{
            Galaxies: jets -- Galaxies: active -- Galaxies: individual: 3C\,84 (NGC\,1275) -- Techniques: interferometric -- Techniques: high angular resolution
               }

   \maketitle

\section{Introduction}\label{sec:Intro}

Jets formed by active galactic nuclei (AGN) are a frequent target of very-long-baseline interferometry (VLBI) observations, with the ultimate goal of shedding light on the physical processes taking place in the ultimate vicinity of supermassive black holes (SMBH).
One method oftentimes employed, is to observe the jetted AGN target at multiple frequencies to draw conclusions about its morphology and spectral properties.

When comparing such different frequency measurements to each other, the effect of the so-called core shift has to be taken into account.
Assuming the standard jet picture put forth in \cite{Blandford79}, a jet is expelled from an optically thick core due to synchrotron self-absorption. 
The size of the core's boundary is frequency dependent; the lower the observing frequency, the further out the boundary extends \citep{Konigl81, Marcaide84, Lobanov98b}.
The apparent radial distance between the boundary at two different observing frequencies is then referred to as the core shift between these frequencies.

Accurately determining the core shift is vital for studying AGN.
For example, studying the spectral index maps of extra-galactic sources heavily relies on the precise alignment of the input images at different frequencies \citep{Lobanov98b}.
A correct image alignment is also a prerequisite to derive a number of physical parameters characterising the jet \citep[e.g.][]{Lobanov98b, Hirotani05, Bach06, Fromm13, Kudryavtseva11, Karamanavis16, Kutkin18, Paraschos21}.

A number of different methods for determining the core shift have been employed in the past successfully.
Initial attempts were made via phase-referencing VLBI observations \citep[see, for example,][]{Marcaide84, Guirado95, Ros01}.
Another procedure, often employed, is aligning optically thin features in jets \citep[e.g.][]{Croke08, Kovalev08, Boccardi16, Pushkarev19}.
A third method, also based on opacity effects, is the measurement of the core size, which is expected to change with frequency \citep[e.g.][]{Kutkin14}.
Finally, a fourth method to align such images of jets utilises the two dimensional cross-correlation, by computing the necessary shift for the cross-correlation function coefficient to reach its peak value \citep[e.g.][]{Fromm13, Paraschos21}.
These methods have proven successful in determining the core shift, however, the first one requires a large amount of allocated resources, the second one necessitates that the core is resolved (which is rarely the case when using ground based VLBI), while the latter two involve observations taken (quasi-simultaneously).
\cite{Pashchenko20}, furthermore, showed that a simple alignment of optically thin features can suffer from biases, resulting in an overestimation of the core shift.

An alternative, more direct observable exists, which is associated with opacity effects and which is not affected by the aforementioned caveats; it is the time lag between the peak of the same flare at different frequencies.
Such time lags between the flare peaks at different frequencies are often seen when studying AGNs \citep[see, for example,][]{Valtaoja92, Bach06, Fromm11, Kudryavtseva11, Kutkin14, Karamanavis16, Kutkin18} and are usually interpreted as the motion of jet features, which move downstream of the bulk jet flow \citep[e.g.][]{Marscher85} and expand adiabatically \citep{vdLaan66}, thus becoming less opaque and allowing radiation to escape.
These time lags can be converted into core shifts if an estimate of the velocity of the features ejected by the jet is available.
It should be noted here, however, that the time lag approach is not free of assumptions, such as the jet flow geometry and jet-feature expansion, and combining it with the aforementioned VLBI approaches yields the most benefit.

A number of approaches have been attempted to extract the time lag information from the total intensity light curves.
For example, \cite{Bach06} used the discrete cross-correlation function method to study the jet of \object{BL\,Lac}.
\cite{Fromm11} used a multi-dimensional $\chi^2$ minimisation of a suit of parameters to model the 2006 flare of the source \object{CTA\,102}.
Almost at the same time, \cite{Kudryavtseva11} introduced a multiple Gaussian function templates procedure to compute the time lag parameters of the source \object{3C\,345}.
\cite{Karamanavis16}, besides using a discrete cross-correlation function approach, also presented a Gaussian process regression method of studying light curves for the source \object{PKS\,1502+106}.

\C\ (\object{NGC\,1275}), which is the focus of this work, is one of the nearest radio galaxies ($D_\textrm{L}= 78.9$\,Mpc, z = 0.0176; \citealt{Strauss92})\footnote{We assume $\Lambda$ cold dark matter cosmology with $H_0 = 67.8$\,km/s/Mpc, $\Omega_\Lambda = 0.692$, and $\Omega_M=0.308$ \citep{Planck16}.}. 
It harbours a SMBH (M$_{\rm BH} \sim 9 \times 10^8$\,M$_\odot$; \citealt{Scharwaechter13}), that serves as the central engine of a powerful, two-sided jet.
Its core and milliarcsecond (mas) region has been imaged in great detail at many frequencies, both in total intensity \citep[e.g.][]{Giovannini18, Paraschos22} and polarisation \citep{Kim19}, revealing a limb-brightened structure \citep[see for example][for a discussion based on MHD simulations]{Kramer21} and multiple features being ejected almost monthly \citep{Paraschos22}.
Locating the exact position of the jet apex might offer an explanation of complex structures observed in the core \citep[e.g.][]{Giovannini18, Kim19, Oh22, Paraschos22}.
Estimates of the jet apex position in \C\ have been presented in the past; \cite{Giovannini18} suggested that the emission north of the 22\,GHz core constitutes the sub-mas counter-jet and thus the jet apex must be within $z_\textrm{core} \sim30\,\mu$as of the 22\,GHz VLBI core.
Using quasi-simultaneous observations of \C\ at 15, 43, and 86\,GHz, and cross-correlating them in two dimensions, \cite{Paraschos21} were able to constrain the location of the jet base to $z_\textrm{core} = 83\pm7\,\mu$as upstream of the 86\,GHz VLBI core.
\cite{Oh22}, on the other hand, used time-averaged 86\,GHz observations, to which they fitted a conical and parabolic expansion profile, which constrained the jet base to $z_\textrm{core} = 55-215\,\mu$as upstream of the 86\,GHz.
In this work, we present the alternative approach of calculating the core shift, and the magnetic field strength which does not rely on imaging \C\ but solely based on available total intensity light curves instead.
We also consider the $\gamma$-ray light curve and its correlation to millimetre radio flux, to constrain possible emission sites.

This paper is structured as follows: in Sect.~\ref{sec:Data+Analysis} we introduce the two methods employed for the further data analysis.
In Sect.~\ref{sec:Results} we present our results, which we then discuss in Sect.~\ref{sec:Discussion}.
Finally, in Sect.~\ref{sec:Conclusions} we summarise our findings.

\section{Data \& Analysis}\label{sec:Data+Analysis}

\begin{figure*}
  \centering
  \includegraphics[scale=0.45]{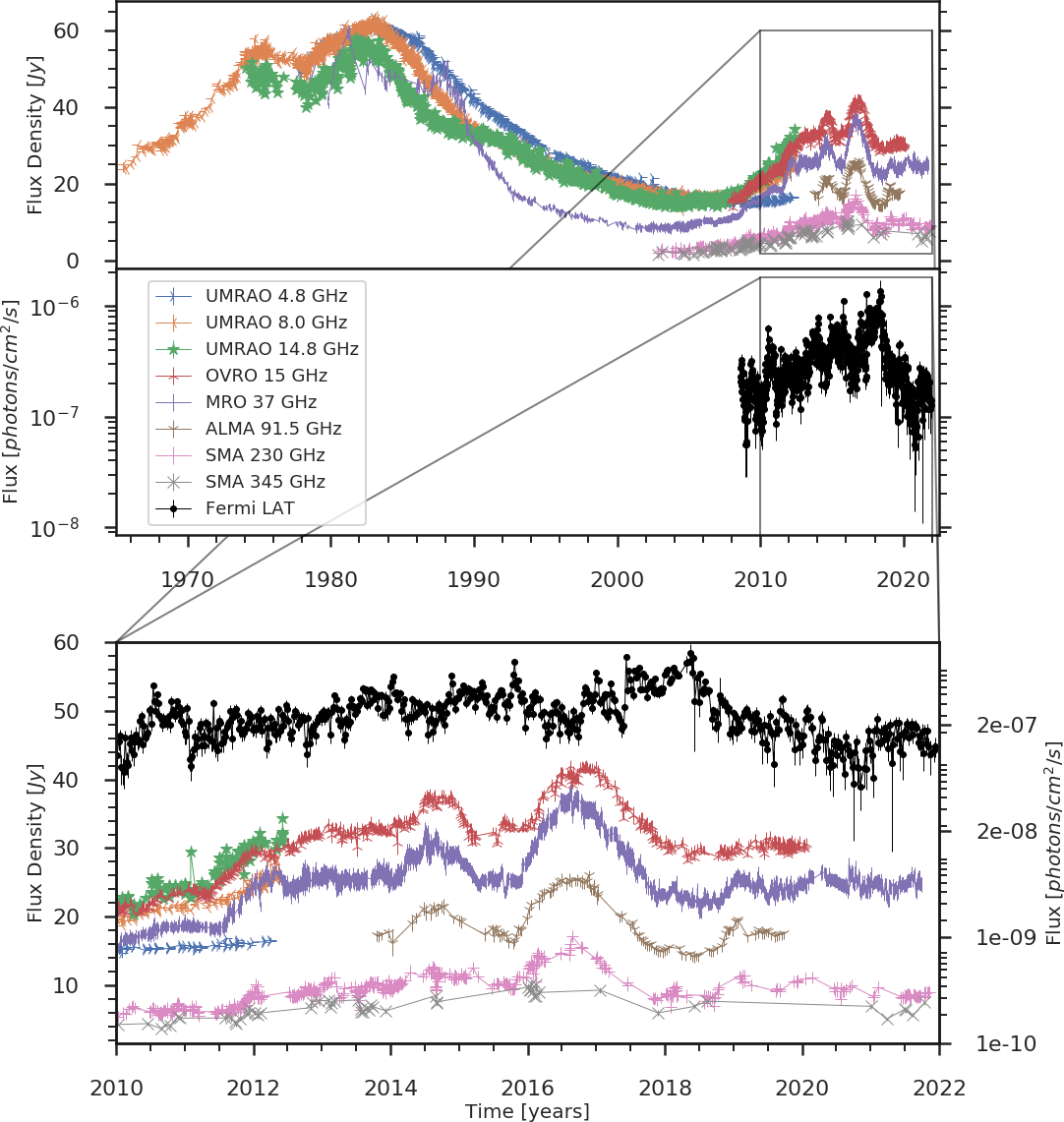}
  \caption{\emph{Top}: Radio light curves of \C, between 1965 and 2022, at 4.8, 8.0, 14.8, 15, 37, 230, and 345\,GHz (coloured entries, in order of appearance in the figure caption).
  \emph{Middle}: The black data set shows the $\gamma$-ray flux.
  \emph{Bottom}: Zoomed-in view of the top two panels to facilitate the comparison of flare onsets and decays in the time range 2010-2022.
  } 
    \label{fig:LCs}
    \end{figure*}

In this section we present the two procedures used to obtain our results.
We employed the total intensity light curve data of \C, at 4.8, 8.0, and 14.8\,GHz (observed at the University of Michigan Radio Observatory; UMRAO), at 15\,GHz (observed at the Owen's Valley Radio Observatory; OVRO, see also \citealt{Richards11}), at 37\,GHz (observed at the Mets\"ahovi Radio Observatory; MRO); at 91.5\,GHz (observed at the Atacama Large Millimeter/submillimeter Array; ALMA), and at 230, and 345\,GHz (observed at the Submillimeter Array; SMA).
The $\gamma$-ray light curve was obtained from the Fermi Large Area Telescope Collaboration\footnote{\url{https://fermi.gsfc.nasa.gov/ssc/data/access/lat/msl_lc/source/NGC_1275}} (Fermi LAT; see also \citealt{Atwood09, Kocevski21}).
This data set (see Fig.~\ref{fig:LCs}) has already been published in \cite{Paraschos22} and we refer the interested reader to that work for further details.
An overview of the light curve parameters is presented in Table~\ref{table:LCparams}.

\begin{table*}[th]
\begin{center}
\begin{threeparttable}
\caption{Light curve parameters.}            
\label{table:LCparams}      
\centering   
\begin{tabular}{cccc}
Observation & Time coverage [yrs] & \# of points & Sampling [\#/yrs]\\
\hline\hline
 4.8 [GHz]  &  1977.8-2012.3 &  1020 &  29.6 \\
 8.0 [GHz]  &  1965.5-2012.3 &  1993 &  42.6 \\
 14.8 [GHz] (UMRAO)   &  1974.1-2012.4 &  1491 &  38.9 \\
 15.0 [GHz] (OVRO)   &  2008.0-2020.1 &  784  &  65.1 \\
 37 [GHz]   &  1979.8-2021.7 &  8907 &  212.4\\
 91.5 [GHz] &  2013.8-2019.7 &  123  &  20.7 \\
 230 [GHz]  &  2002.8-2021.9 &  750  &  39.4 \\
 345 [GHz]  &  2002.9-2021.8 &  116  &  6.1  \\
 $\gamma$   &  2008.6-2021.9 &  680  &  51.0 \\
\hline
\end{tabular}
\end{threeparttable}
\end{center}
\end{table*}

\subsection{Gaussian process regression}\label{ssec:GPR}

The first implementation chosen to compute the time lags between the light curves and, by extension, the core shifts is that of the Gaussian process regression (GPR) \citep{Rasmussen06}.
GPR has been increasingly used in astronomy \citep{Karamanavis16, Kutkin18, Mertens18, Pushkarev19}, as it takes advantage of the benefits of machine learning.
Specifically, GPR offers a non-parametric approach to data fitting, which does not require any assumptions about the fitting function.
Instead, a set of so-called hyperparameters is used to draw functions defined by a kernel (i.e. the covariance between two function values).
In this work, we will refrain from presenting a detailed description of GPR, as this has already been done in past works \citep[e.g.][]{Karamanavis16}; rather we will directly describe our methodology in brief.

Since our data set consists of multiple observations of \C, taken at different frequencies, over many decades, the choice of a suitable kernel has to be carefully motivated. 
A simple square exponential kernel \citep{Karamanavis16} is suited for data sets with a single characteristic time scale.
As is shown in Fig.~\ref{fig:LCs}, however, \C\ exhibits many flares over the years, of varying duration.
Hence, we employed a more generalised kernel, which is the sum of multiple square exponential kernels, which is called rational quadratic kernel and is given by the following equation:
\begin{equation}
    k_\textrm{RQ}\left(t_\textrm{m}, t_\textrm{n}\right) = N^2\left(1+\frac{\left(t_\textrm{m} - t_\textrm{n}\right)^2}{\left(2\alpha l^2\right)}\right)^{-\alpha},\label{eq:rq}
\end{equation}
where $\boldsymbol{\theta} = \{N, l, \alpha\}$ is the hyperparamater vector of the kernel $k_\textrm{RQ}$, $N$ is the amplitude scaling, $l$ is the characteristic time scale and $\alpha$ specifies the weighting of the different such time scales.
Following \cite{Kutkin18}, and in order to model the noise, we added the white kernel
\begin{equation}
    k_\textrm{W} = \delta_\textrm{mn}\left(\sigma^2\right), \label{eq:w}
\end{equation}
where $\delta_\textrm{mn}$ is the Kronecker delta and $\sigma^2 = \sigma^2_\textrm{obs} + \sigma^2_\textrm{jit}$.
Here $\sigma_\textrm{obs}$ is the uncertainty inherent to the measurements of the data and $\sigma_\textrm{jit}$ is possibly unaccounted for noise manifested as jitter.
Combining Eqs.~\ref{eq:rq} and \ref{eq:w}, we arrived at the kernel used in this work for the GPR:
\begin{equation}
k_\textrm{tot} = k_\textrm{RQ} + k_\textrm{w}. \label{eq:tot}    
\end{equation}

\begin{figure*}
  \centering
  \includegraphics[scale=0.35]{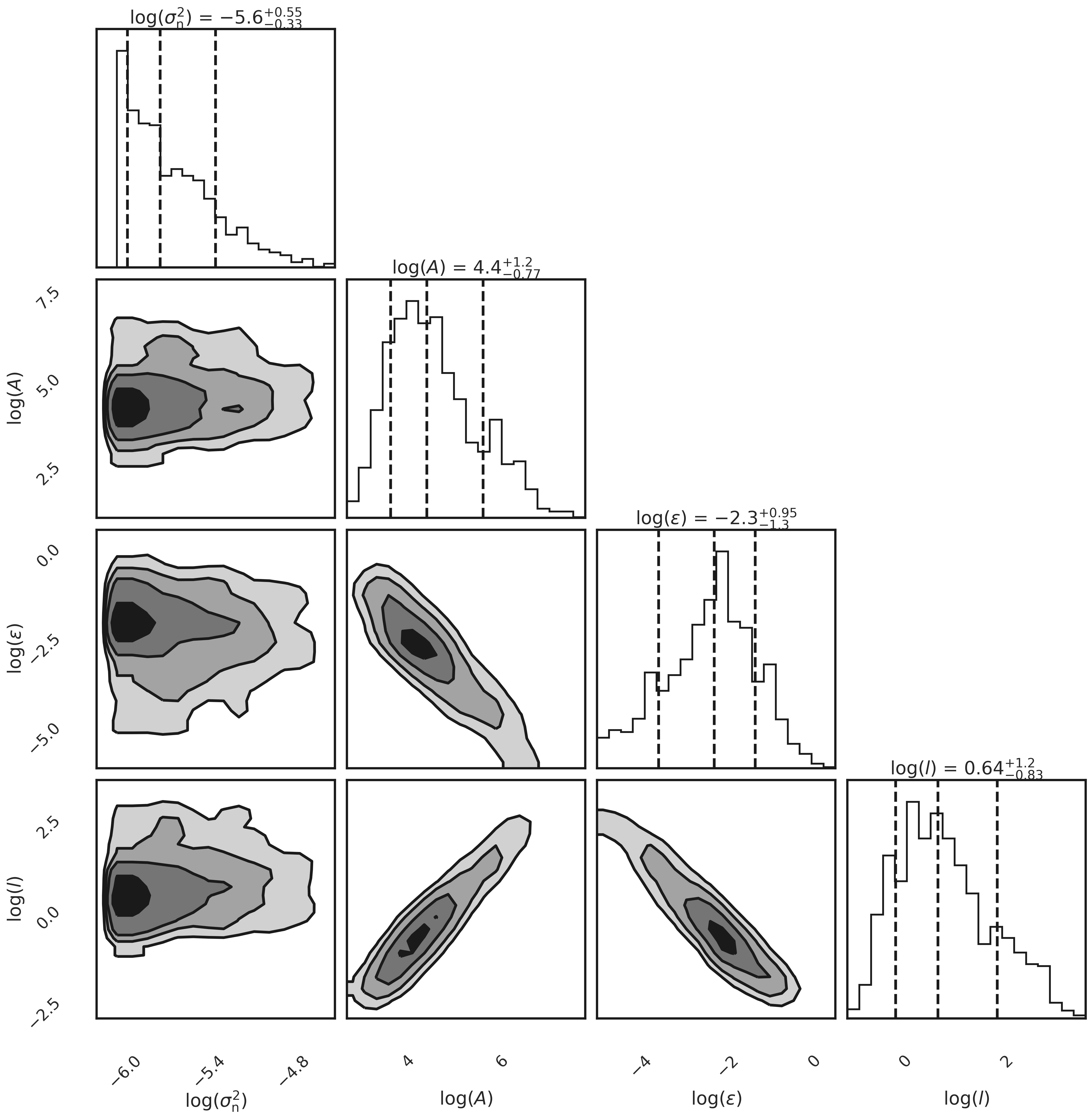}
  \caption{Exemplary posterior distributions of the GPR hyperparameters at 15\,GHz (OVRO).
  The dashed lines correspond to the 16, 50, and 84 percentiles.
  } 
    \label{fig:corner}
    \end{figure*}

In order to apply the GPR to \C, we first subtracted the minimum flux per frequency $S_0$ (see Table~\ref{table:Synth}), to account for the possibility that the flux does not reach a quiescent state between consecutive flares \citep{Karamanavis16, Kutkin18}.
To estimate the uncertainties of the inferred hyperparameters, we sampled their posterior distributions via a Markov chain Monte Carlo (MCMC) method.
We used the \textsc{emcee} sampler by \cite{Foreman13} and the \textsc{george} Python package \citep{Ambikasaran15}, and implemented the description of the hyperparameter vector construction presented in \cite{Kutkin18}.
As an example\footnote{Produced with the \textsc{corner} package \citep{Foreman21}.}, we show in Fig.~\ref{fig:corner} the posterior distributions of the hyperparameters of the 15\,GHz OVRO observations.
The resulting posterior means and uncertainty bands are presented in Fig.~\ref{fig:GP}.
From these, the exact flaring times were then extracted and the time lag $\tau_\textrm{GP}$ computed (see Table~\ref{table:GPR-DCF}).

\subsection{Discrete cross-correlation function}\label{ssec:DCF}

A more standard method of evaluating characteristic variability time scales in light curves \citep[see e.g.][for a discussion of the variability in a large sample of AGNs including \C]{Hovatta07}, as well as time lags between light curves \citep[e.g.][]{Kutkin14, Fuhrmann14, Rani17, Hodgson18} is via the discrete cross-correlation function \citep[DCF;][]{Edelson88}.
The DCF is a widely used procedure to find correlations between two data sets, as it yields robust results even for unevenly sampled data, with known measurement errors.
Another advantage of this method is that it does not require the input data sets to be of the same length.
It also avoids interpolating in the time domain, while also providing error estimates.
A positive time lag between two light curves $LC_\textrm{i}$ and $LC_\textrm{j}$ (with $\nu_\textrm{i}>\nu_\textrm{j}$) indicates that features in $LC_\textrm{i}$ are leading those in $LC_\textrm{j}$.
A negative time lag on the other hand indicates the opposite.
For the work presented here, we used the publicly available implementation presented in \cite{Robertson15}.
The determination of the confidence bands is presented in Appendix~\ref{App:DCF_Parameters}.
The results are presented in Fig.~\ref{fig:DCF} and the time lag values $\tau_\textrm{DCF}$ in Table~\ref{table:GPR-DCF}.
A discussion of the uncertainties estimate is provided in Appendix~\ref{App:DCF_Parameters}.

\section{Results}\label{sec:Results}

Equipped with the tools presented in Sect.~\ref{sec:Data+Analysis}, we present here our results.
Details of the implementations are discussed in Appendix~\ref{App:GPR} and \ref{App:DCF}.

\subsection{Time lags}\label{ssec:timelags}

\begin{figure*}
  \centering
  \includegraphics[scale=0.2]{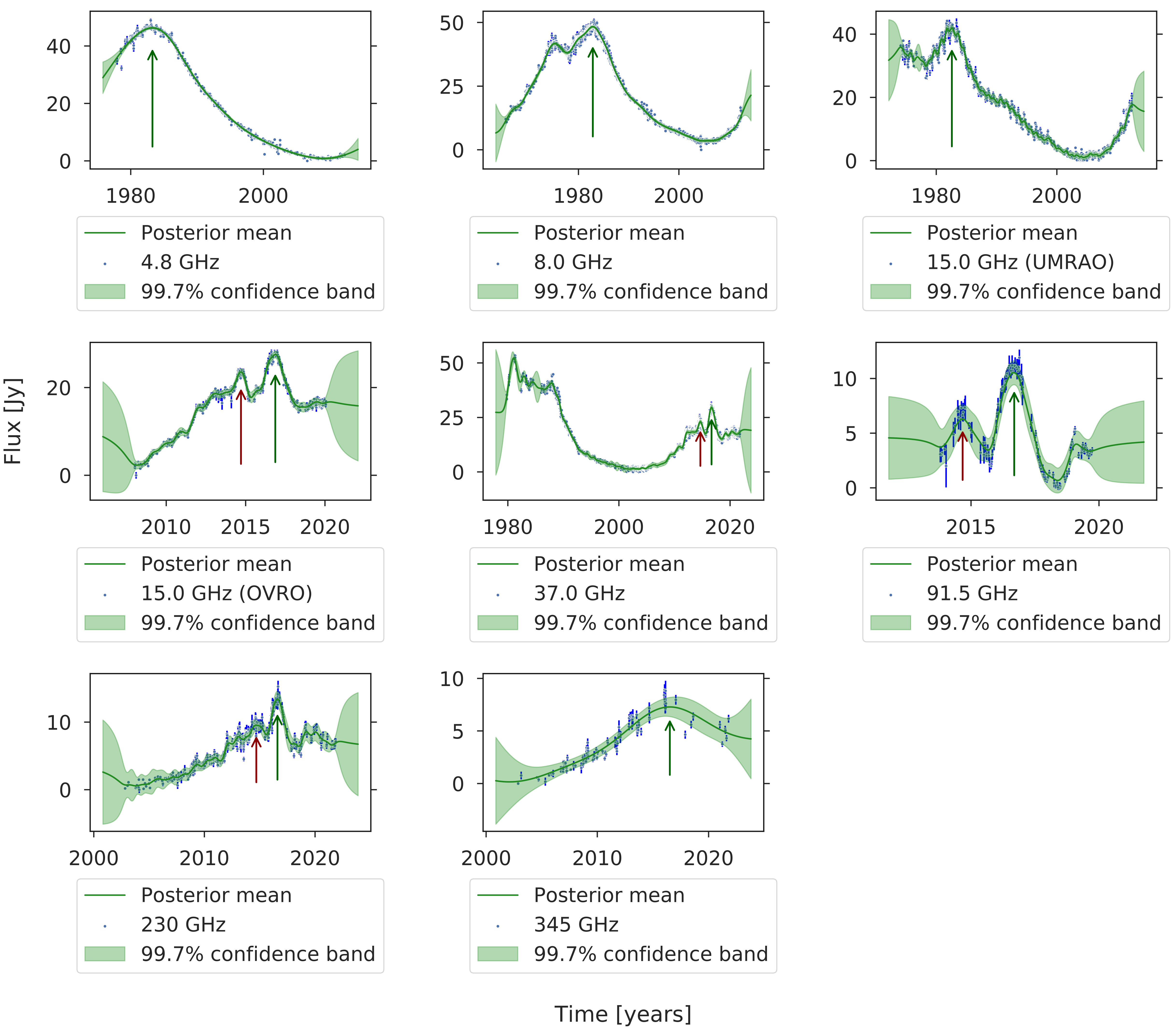}
  \caption{GPR curves of \C, using the light curves presented in Fig.~\ref{fig:LCs}.
  The minimum flux density shown in Table~\ref{table:Synth} has already been subtracted.
  The blue data points denote the observations, the green solid line the posterior mean, and the light-green envelope shows the 99.7\% confidence bands.
  The dark-green arrow denotes the primary flare per frequency used for the time lag calculation, and the dark-red arrow the secondary flare, wherever it was detectable.
  } 
    \label{fig:GP}
    \end{figure*}

\begin{figure*}
  \centering
  \includegraphics[scale=0.2]{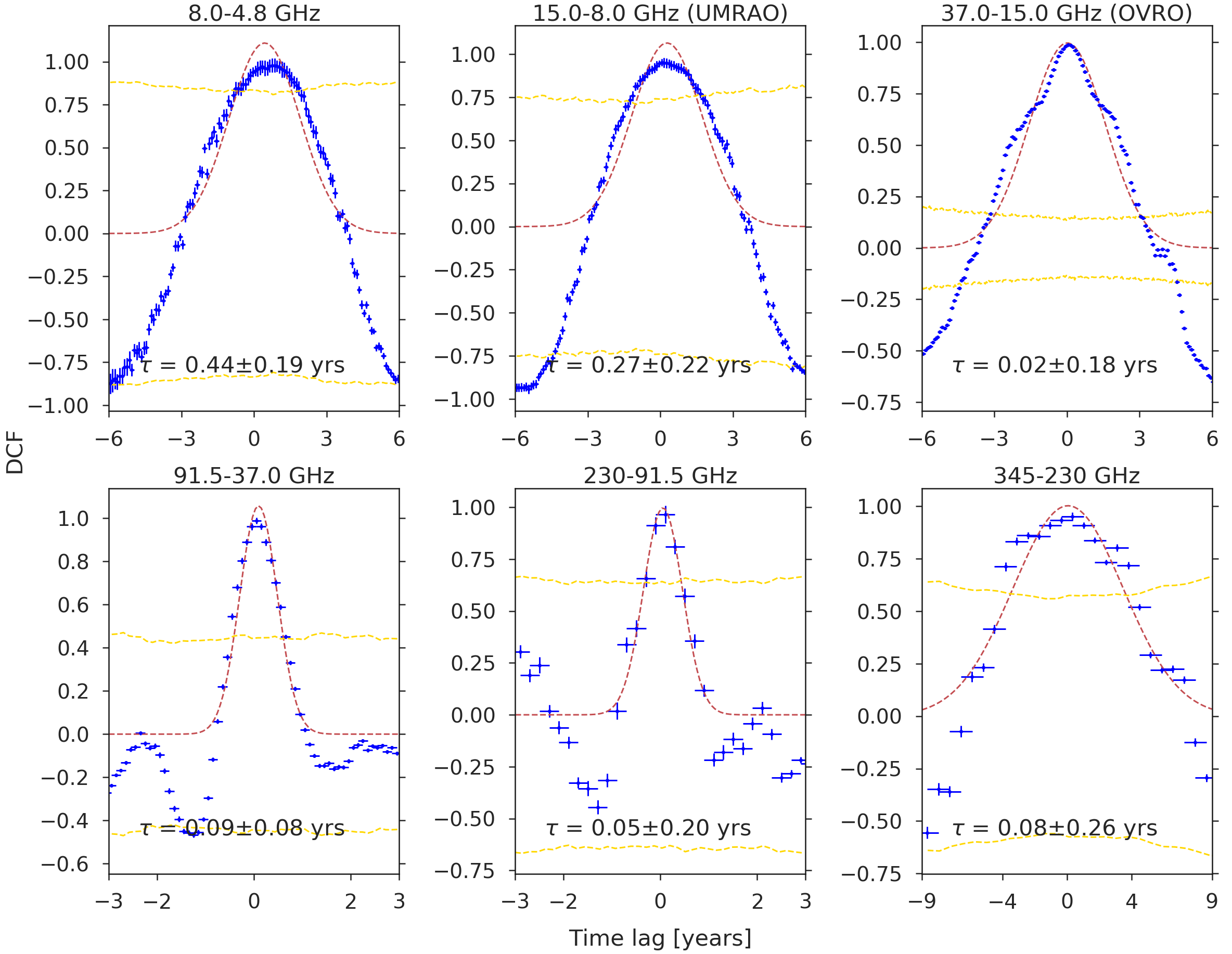}
  \caption{DCF analysis curves. \emph{Top row}: 8.0\,GHz versus 4.8\,GHz, 15.0\,GHz (UMRAO) versus 8.0\,GHz, and 37.0\,GHz versus 15.0\,GHz (OVRO).
  \emph{Bottom row}: 91.5\,GHz versus 37\,GHz, 230\,GHz versus 91.5\,GHz, and 345\,GHz versus 230\,GHz.
  The dashed red line corresponds to a Gaussian function fit of the DCF, the mean of which ($\tau$) is used as a more accurate estimate of the peak position.
  The dashed gold curve denotes the 99.7\% confidence band.
  } 
    \label{fig:DCF}
    \end{figure*}

Our results, using both GPR and DCF approaches are shown in Figs.~\ref{fig:GP} and \ref{fig:DCF}, and summarised in Table~\ref{table:GPR-DCF}.
We find that the flares consistently appear first at higher frequencies and later at lower ones.
The very good agreement between both approaches allowed us to average the time lags and compute an average time lag $\langle\tau\rangle$ per frequency to the reference frequency 345\,GHz.
To compute the time lag between a given frequency and the reference frequency, we calculated the cumulative sum of the time lags between them \citep[see also][]{Karamanavis16}.
For example, $\Delta t_\textrm{DCF}^\textrm{91.5-345\, GHz} = \Delta t_\textrm{DCF}^\textrm{91.5-230\,GHz} + \Delta t_\textrm{DCF}^\textrm{230-345\,GHz}$. 
Furthermore, the DCF analysis shows that all examined radio light curve pairs are significantly correlated with each other.

\begin{figure}
  \centering
  \includegraphics[scale=0.35]{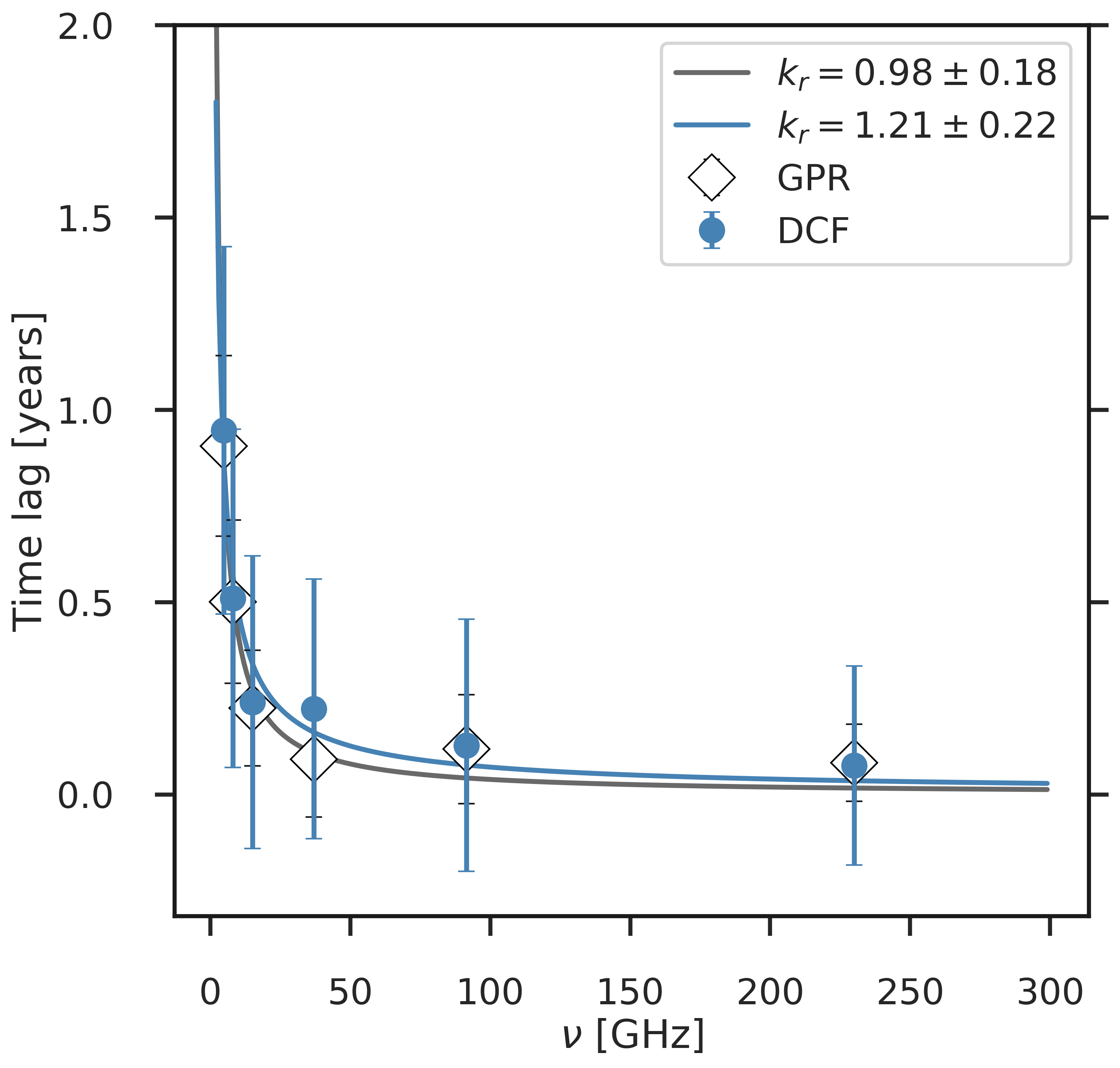}
  \caption{Temporal core offset position fit, with respect to the reference frequency (345\,GHz).
  The open data points and grey curve denote the GPR values, and the blue data points and curve show the DCF values.
  } 
    \label{fig:Fit_aver}
    \end{figure}

Figure~\ref{fig:Fit_aver} displays the values of Table~\ref{table:GPR-DCF}, which evidently follow a power law of the form $\Delta t \propto \nu^{-1/k_\textrm{r}}$.
Under the assumption that the size of the emission region increases linearly with the distance $r$ from the core ($z_\textrm{core}\propto r$), the magnetic field strength and particle density decrease with distance as $B \propto r^{-b}$ and $N \propto r^{-n}$
\citep{Lobanov98a, Fromm13, Paraschos21}; then $k_\textrm{r}$ is connected to $n$ (number density power law index), $b$ (magnetic field power law index), and the optically thin spectral index $\alpha_\textrm{thin}$ via the following relation:
\begin{equation}
    k_\textrm{r} = [2n + b(3 - 2\alpha_\textrm{thin} ) - 2]/(5 - 2\alpha_\textrm{thin} ).\label{eq:k}
\end{equation}
We fitted a relation of this form to the time lags from both methods (using a least squares minimisation approach) and found that $k_\textrm{r}^\textrm{GPR} = \kgpr$, and $k_\textrm{r}^\textrm{DCF} = \kdcf$.
Fitting the average of the two methods produces an average power law index value of $k_\textrm{r}^\textrm{aver} = \kav$.
A value of $k_\textrm{r} = 1$ implies that there is equipartition between the particle energy and magnetic field energy, in which case it is usually assumed that $b=1$ and $n=2$ \citep{Lobanov98b}.

\begin{threeparttable}[h]
\caption{Results of the GPR and DCF analyses for the light curves of \C.}    
\label{table:GPR-DCF}     
\centering    
\begin{tabular}{cccc}
$\nu$ [GHz] & $\Delta\tau_\textrm{GPR}$ [yrs] & $\Delta\tau_\textrm{DCF}$ [yrs] & $\langle\tau\rangle$ [yrs]\\
\hline\hline
 4.8   &  0.91$\,\pm\,$0.23 &  0.95\,$\pm$\,0.48 &  0.93\,$\pm$\,0.38\\
 8.0   &  0.50$\,\pm\,$0.21 &  0.51\,$\pm$\,0.44 &  0.51\,$\pm$\,0.35\\
 15.0  &  0.22$\,\pm\,$0.15 &  0.24\,$\pm$\,0.38 &  0.23\,$\pm$\,0.29\\
 37.0  &  0.09$\,\pm\,$0.15 &  0.22\,$\pm$\,0.34 &  0.16\,$\pm$\,0.26\\
 91.5  &  0.12$\,\pm\,$0.14 &  0.13\,$\pm$\,0.33 &  0.12\,$\pm$\,0.25\\
 230.0 &  0.08$\,\pm\,$0.10 &  0.08\,$\pm$\,0.26 &  0.08\,$\pm$\,0.20\\
\hline
\end{tabular}

\begin{tablenotes}
  \small
  \item \textbf{Notes:} Columns: (1) observing frequency; (2) time lag of observing frequency to the reference frequency (345\,GHz), computed via GPR; (3) time lag of observing frequency to the reference frequency (345\,GHz), computed via DCF; (4) average of time lags $\tau_\textrm{GPR}$ and $\tau_\textrm{DCF}$.
\end{tablenotes}
\end{threeparttable}

\subsection{Core shift}\label{ssec:coreshift}

Transforming time lags into core shift measurements can be done using the standard Eqs.~(4) and (7) presented in \cite{Lobanov98a} and \cite{Hirotani05} \citep[see also][]{Paraschos21}.
$\Omega_\textrm{r}^{\nu}$ is the measure of the core position offset and in our case can be calculated by replacing $\Delta r_\textrm{obs}$ with $\mu\langle\tau\rangle$ \citep[see, for example,][]{Kudryavtseva11}.
Here $\mu$ is the apparent component velocity in mas/yr, given by the following relation:
\begin{equation}
    \mu = \frac{c\left(1+z\right)\beta_\textrm{app}}{D_\textrm{L}}, \label{eq:mu}
\end{equation}
with $c$ the velocity of light, $z$ the redshift, $\beta_\textrm{app}$ the apparent velocity in units of $c$, and $D_\textrm{L}$ the luminosity distance.
For the viewing angle $\theta$ we adopted the values used in \cite{Paraschos21}.
For $\beta_\textrm{app}$ we used the upper and lower velocities found by \cite{Paraschos22}.
These parameters are shown in Table~\ref{table:Parameters}.

\begin{threeparttable}[h]
\caption{Parameters used in this work.}             
\label{table:Parameters}     
\centering    
\begin{tabular}{cc}
Parameters & Values\\
\hline\hline
$\theta$ & $20^\circ-65^\circ {}^{[a]}$\\
$P_\textrm{jet}$ & $10^{44}-10^{45} \textrm{erg s}^{-1} {}^{[a,d,e]}$\\
$\alpha_\textrm{thin}$ & $-0.77^{[a]}$\\
$\gamma_\textrm{max}$ & $10^3-10^5 {}^{[a]}$\\
$\gamma_\textrm{min}$ & $1^{[a]}$\\
$\phi$ & $2.8^\circ-20^\circ {}^{[a]}$\\
\hline
$\beta_\textrm{app}$ & $0.03-0.12^{[b]}$\\
\hline
$\xi$ & $0.01^{[c]}$\\
\hline
\end{tabular}

\begin{tablenotes}
  \small
  \item $^a$\cite{Paraschos21}; $^b$\cite{Paraschos22}; $^c$\cite{Sironi13}; $^d$\cite{Abdo09}; $^e$\cite{Magic18}
\end{tablenotes}
\end{threeparttable}\\

Furthermore, for the determination of the core shift values, presented in Table~\ref{table:Output}, we used the averaged power law index $k_\textrm{r}^\textrm{aver}=\kav$.
A clear trend is confirmed; the lower the frequency of the observed VLBI core, the farther away it is situated from the 345\,GHz VLBI core used as reference.
Under the assumption that the distance between the jet base and the 345\,GHz VLBI core is negligible, we can compare the distance of the 91.5\,GHz VLBI core calculated here, to the one determined in \cite{Paraschos21} and \cite{Oh22} (at 86\,GHz).
In particular, we find the distance of the 91.5\,GHz VLBI core to the jet base to be between $z_\textrm{91.5\,GHz}=\znf$.
This value range overlaps with the lower estimates found by \cite{Paraschos21} ($z_\textrm{86\,GHz}=400-1500\,\textrm{R}_\textrm{s}$) and \cite{Oh22} ($z_\textrm{86\,GHz}=200-3000\,\textrm{R}_\textrm{s}$; adjusted to the same BH mass and viewing angle assumptions).
It should be pointed out here, that the results discussed above assume a homogeneous and conical jet geometry, as in \cite{Blandford79}.
Should the core region become optically thin just above 86\,GHz, our results and relevant discussions might be affected.

\begin{table*}[th]
\begin{center}
\begin{threeparttable}
\caption{Core shift and magnetic field of \C, with respect to the reference frequency 345\,GHz.}            
\label{table:Output}      
\centering    
\begin{tabular}{ccccc}
$\nu$ [GHz] & $\Omega_{r\nu}$ [pc GHz$^{1/k_\textrm{r}}$] & $r_\textrm{core}$ [$10^{-3}$\,pc] ([$\textrm{R}_\textrm{s}$]) & B$_\textrm{core}$ [G] & B$_\textrm{1pc}$ [mG]\\
\hline\hline
 4.8   &  0.04-0.14 &  9.41-99.70 (120.62-3387.81) &  0.24-0.69  &  4.06-37.42  \\
 8.0   &  0.03-0.13 &  5.19-55.04 (66.58-1870.11)  &  0.40-1.17  &  3.81-35.11  \\
 15.0  &  0.03-0.11 &  2.45-25.95 (31.40-881.85)   &  0.77-2.26  &  3.47-31.95  \\
 37.0  &  0.05-0.18 &  1.79-18.98 (22.96-644.89)   &  1.66-4.85  &  5.45-50.19  \\
 91.5  &  0.10-0.40 &  1.73-18.34 (22.18-623.11)   &  3.34-9.78  &  10.62-97.80 \\
 230.0 &  0.34-1.37 &  2.52-26.70 (32.30-907.34)   &  6.20-18.14 &  28.68-264.17\\
\hline
\end{tabular}
\begin{tablenotes}
  \small
  \item \textbf{Notes:} Columns: (1) observing frequency; (2) core position offset; (3) distance between the frequency-dependent radio core and the jet base (projected in pc and de-projected in $\textrm{R}_\textrm{s}$); (4) magnetic field strength at the radio core; (5) magnetic field strength at a distance of 1\,pc from the jet base.
\end{tablenotes}
\end{threeparttable}
\end{center}
\end{table*}

\subsection{Magnetic field estimation}\label{ssec:Bfield}

Our measurement of $k_\textrm{r}^\textrm{aver}=\kav$ indicates, that the system is in equipartition within the error budget and we can thus use Eqs. (B.2) and (B.3) from \cite{Paraschos21} to compute the equipartition magnetic field \citep[see also][]{Hirotani05} at the core and at a distance of 1\,pc (see Table~\ref{table:Output}).
The magnetic field strength ranges between $B_\textrm{core}^{4.8\,\textrm{GHz}} \approx 0.46$\,G and  $B_\textrm{core}^{230\,\textrm{GHz}} \approx 13.3$\,G.
The assumed values for the optically thin spectral index $\alpha_\textrm{thin}$, the maximum and minimum Lorentz-factors $\gamma_\textrm{max}$ and $\gamma_\textrm{min}$, as well as the jet half opening-angle are listed in Table~\ref{table:Parameters}.
Our findings for the magnetic field strength at 91.5\,GHz ($B_\textrm{core}^\textrm{91.5\,GHz} = \Bno$\,G) are in good agreement with the value range at 86\,GHz presented in \cite{Paraschos21} ($B_\textrm{core}^{86\,\textrm{GHz}} = 1.8-4.0$\,G).
Our result is slightly lower than the synchrotron self-absorption magnetic field calculated by \cite{Kim19} ($B_\textrm{SSA}\sim21\pm14$\,G) but still in agreement within the error budget.

\section{Discussion}\label{sec:Discussion}

\subsection{Emission site of $\gamma$-rays}\label{ssec:gamma}

The answer to the question whether $\gamma$-rays are predominantly emitted upstream or downstream of the radio emission is one of active debate.
A number of works \citep[e.g.][]{Jorstad01, Lahteenmaki03, LeonTavares11, Fuhrmann14} suggest that the radio flux is generally simultaneous or trailing the $\gamma$-rays.
Furthermore, \cite{Pushkarev10} and \cite{Kramarenko22} found that the 15\,GHz VLBA emission generally lags behind the $\gamma$-ray emission.
At a slightly higher frequency (37\,GHz) \cite{Ramakrishnan15} and \cite{Ramakrishnan16} found that in most sources with a correlation between radio flux and $\gamma$-rays, the $\gamma$-rays lead the radio flux and the start times of the radio and $\gamma$-ray activity correlate.
It is interesting to note here that \cite{Ramakrishnan15} found in their sample two sources (\object{1633+382} and \object{3C\,345}), where the $\gamma$-rays trail the radio emission.
Below we attempt to answer this question for \C.

  \begin{figure}
  \centering
  \includegraphics[scale=0.2]{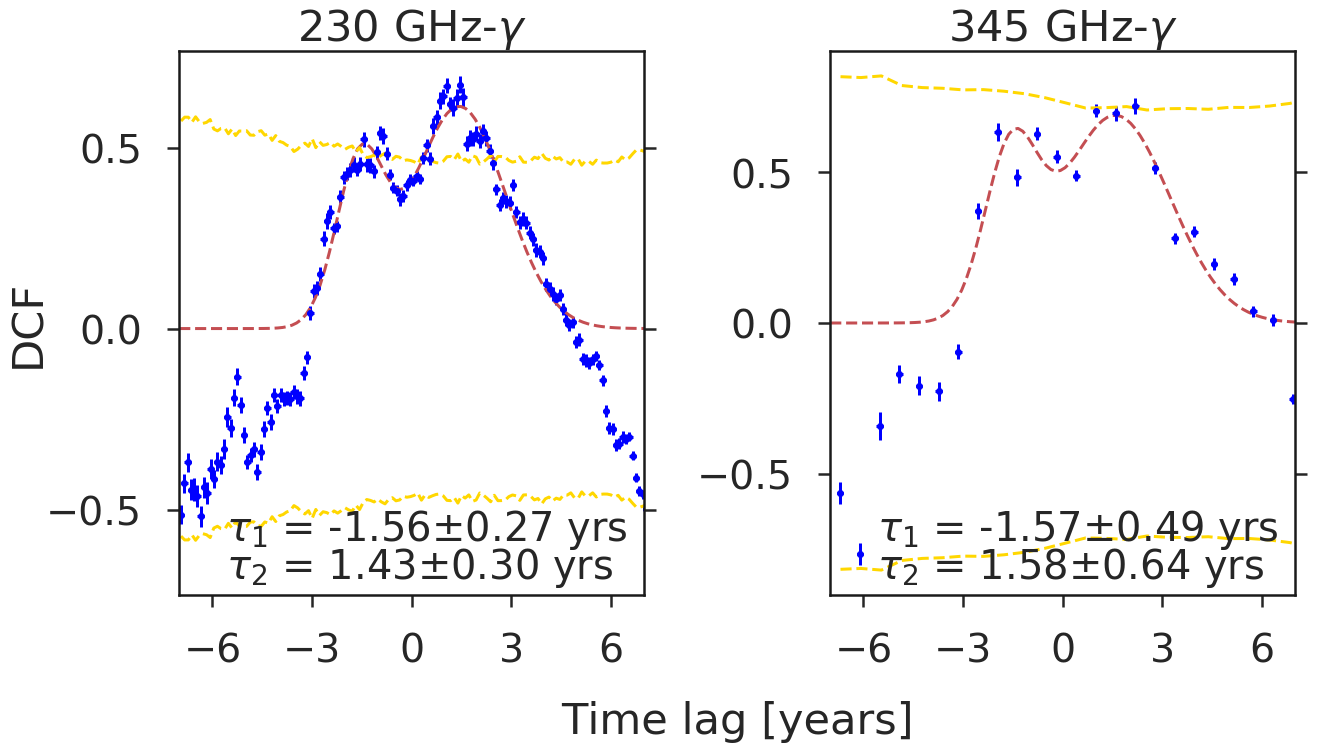}\hfill
  \includegraphics[scale=0.19]{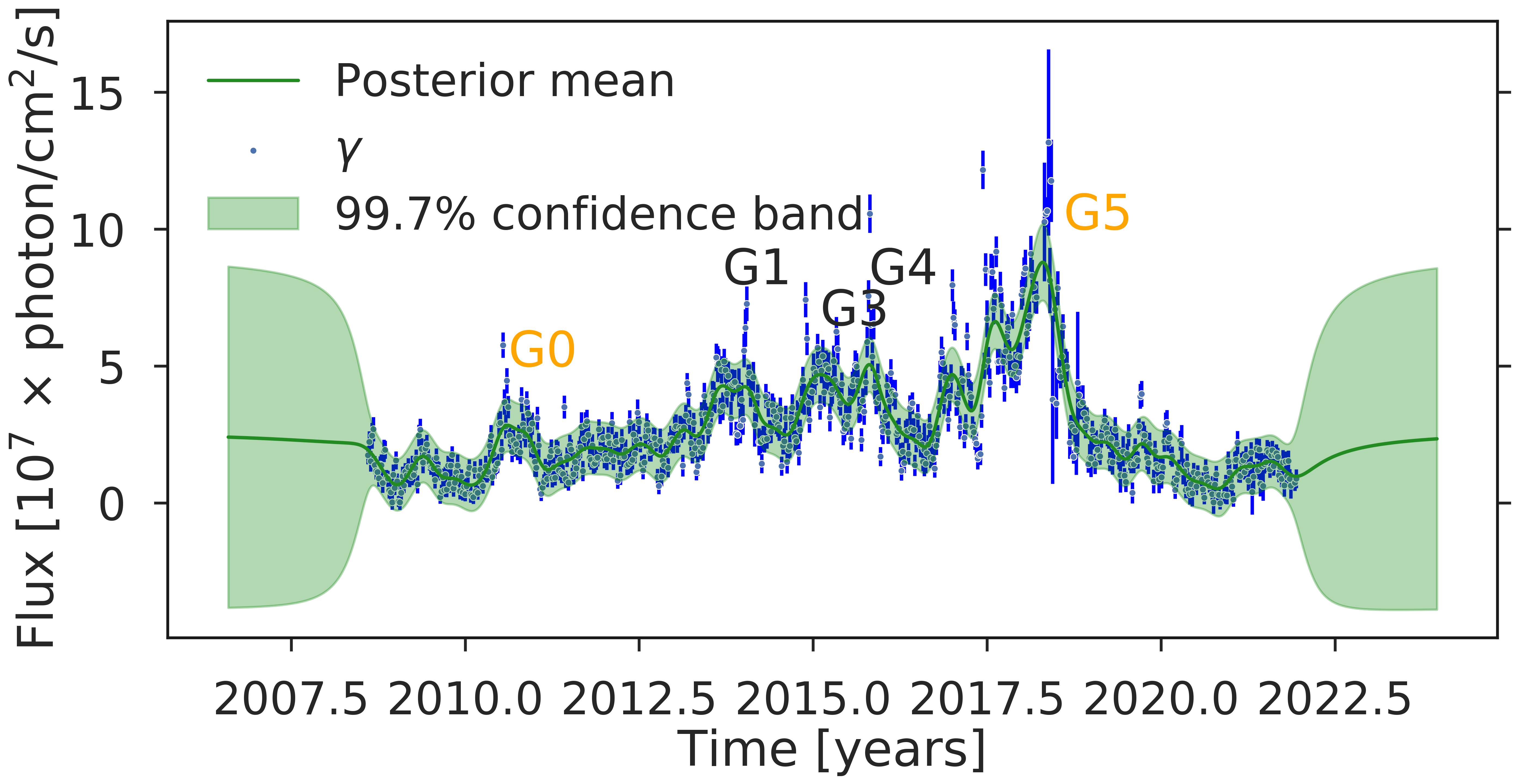}\hfill
  \includegraphics[scale=0.19]{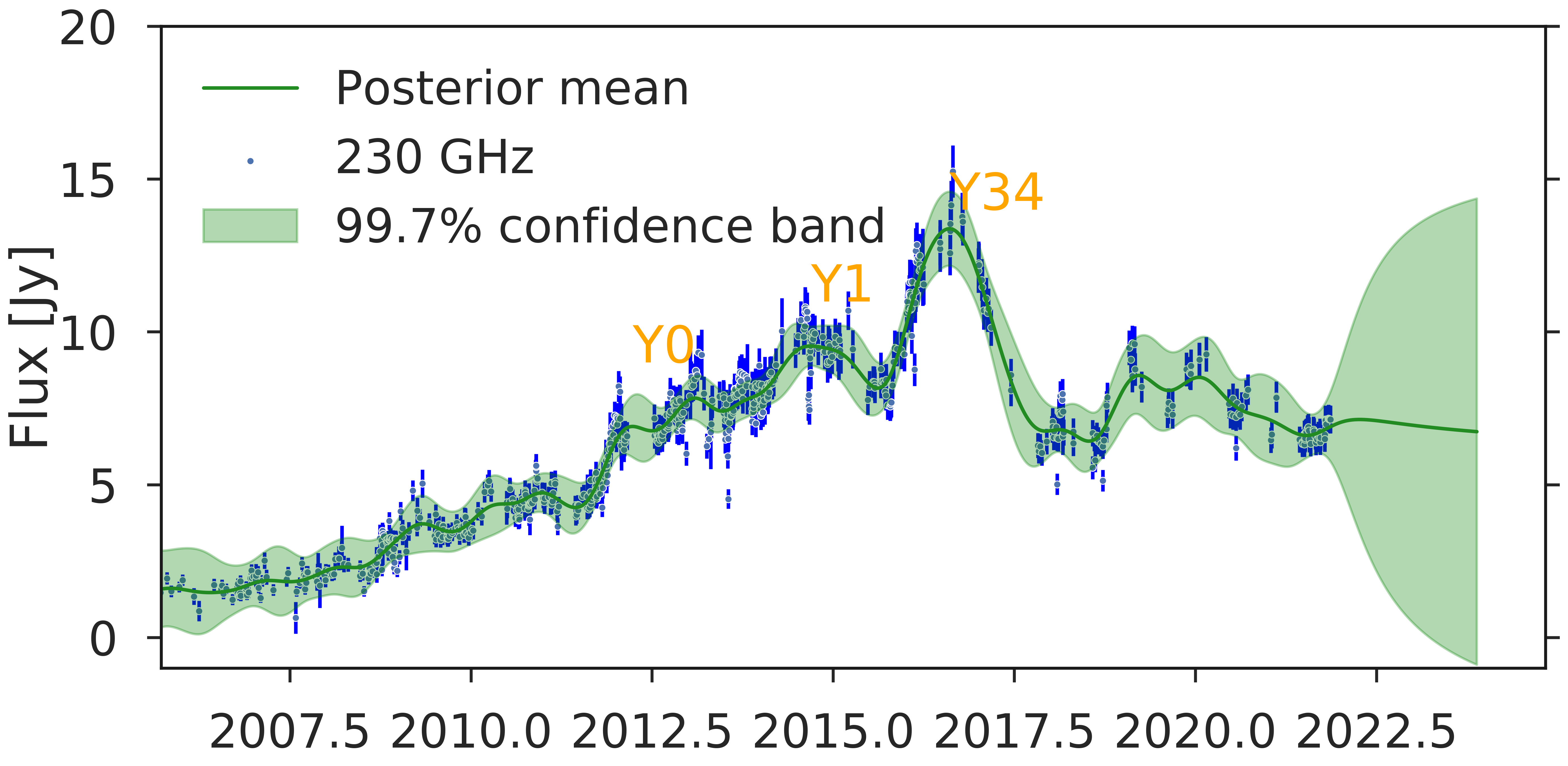}
  \caption{
  DCF and GPR analyses of the $\gamma$-ray light curve.
  \emph{Top row}: DCF of $\gamma$-rays versus 230\,GHz (left) and 345\,GHz (right) for \C.
  The dashed red line corresponds to a double Gaussian fit of the DCF, the means of which ($\tau_1$ and $\tau_2$) are used as a more accurate estimate of the peak positions.
  The dashed gold curve again denotes the 99.7\% confidence band.
  \emph{Middle row}: GPR curve for $\gamma$-ray emission in \C.
  G1, G3, G4 \citep{Hodgson18}, and G0, G5 denote the $\gamma$-ray flares considered in this work.
  \emph{Bottom row}: GPR curve for 230\,GHz emission in \C, matching the time range of the $\gamma$-rays.
  Y0, Y1, and Y34 denote the most prominent flares, which are possibly associated with the $\gamma$-ray ones.
  } 
    \label{fig:DCF_gamma}
    \end{figure}

The $\gamma$-ray activity in \C\ can be split into short term outbursts on top of a long term, slowly increasing trend \citep[also seen in the optical V band by][]{Nesterov95}.
It is thought that the latter most likely originates in the C3 region \citep{Nagai14}, which is a bright feature at a distance of $\sim1$\,pc from the VLBI core \citep{Dutson14, Nagai16, Hodgson18, Linhoff21}.

The short term outburst location, on the other hand, is ambiguous.
\cite{Nagai16} attributed the lack of correspondence between $\gamma$-ray emission and radio flux to the $\gamma$-rays being emitted in the spine \citep{Ghisellini05} of the jet, corroborated by the detected limb-brightened jet structure of \C\ \citep{Nagai14, Giovannini18}.
\cite{Hodgson18} postulated that, if the time lag between $\gamma$-ray emission and 1\,mm (230\,GHz) radio emission is simply due to opacity effects, the emission region would be 0.07\,pc ($900-2400\,\textrm{R}_\textrm{s}$) upstream of the 1\,mm (230\,GHz) radio core.
On the other hand, \cite{Britzen19} found that the $\gamma$-rays might be emitted downstream of the radio emission, by using $\gamma$-ray data up to $\sim2018.5$.
\cite{Linhoff21} determined that the emission region lies outside the Ly $\alpha$ radius and perhaps even beyond the broad line region (BLR; $\textrm{R}_\textrm{BLR}\sim6.5-9.1\times10^{-3}$\,pc, corresponding to $80-310\,\textrm{R}_\textrm{s}$; see \citealt{Oh22}).
Most recently, \cite{Hodgson21} attempted to connect the kinematics of the jet of \C\ to the $\gamma$-ray emission, using the WISE package \citep{Mertens15}.

In our study, we expand on previous works by including all available Fermi-LAT $\gamma$-ray data, up to 2021.9.
As shown in Fig.~\ref{fig:DCF_gamma}, we find that the $\gamma$-ray emission is significantly correlated with the 1\,mm (230\,GHz) radio flux, and rather marginally with 0.8\,mm (345\,GHz) one.
Specifically, we find that the $\gamma$-rays either precede the 230\,GHz flux (345\,GHz flux) by $\tau_{\gamma-\textrm{230\,GHz}} =\tauonemmgammap$\,years ($\tau_{\gamma-\textrm{345\,GHz}}= \taupemmgammap$\,years) or trail the 230\,GHz flux (345\,GHz flux) by $\tau_{\textrm{230\,GHz-}\gamma}= \tauonemmgammat$\,years ($\tau_{\textrm{345\,GHz-}\gamma}= \taupemmgammat$\,years).
These values are higher than the ones reported by \cite{Hodgson18}, whose $\gamma$-ray data set spanned until <2017.
The positive time lag compares well with the 1.37\,years reported in the work by \cite{Ramakrishnan15}, in which the authors cross-correlated the $\gamma$-rays with the 37\,GHz flux for the time period 2008.6-2013.6.
It should be pointed out here, that the $\gamma$-ray light curve's sampling rate is higher than the one of the 230\,GHz (345\,GHz) flux (see Table~\ref{table:LCparams}) by $\sim20\%$ ($\sim90\%$).
Significantly lower sampling rates at higher frequencies, as is the case for the 345\,GHz -- $\gamma$-ray cross-correlation analysis, might result in missing the flare peaks of rapid flares.
In our analysis, however, there is good agreement between the DCF time lags of the 345\,GHz -- $\gamma$-ray pair and the 230\,GHz -- $\gamma$-ray pair, as well as overall consistency between the DCF and GPR method results.

Under the same assumptions used to determine the core shifts in Sect. \ref{ssec:coreshift}, we find that the $\gamma$-ray emission site would be $z_\gamma=\zonemm\,\textrm{R}_\textrm{s}$ ($z_\gamma=\zpemm\,\textrm{R}_\textrm{s}$) of the 230\,GHz (345\,GHz) VLBI core.
Comparing these values to the distance between the 230\,GHz VLBI core and the jet base (see Table~\ref{table:Output}), we find that the $\gamma$-ray emission region would be close to or inside the BLR,which has been deemed unlikely, as discussed above, by the work presented in \cite{Linhoff21}.
Assuming again that the distance between the 345\,GHz radio core and jet apex and BH is negligible, the $\gamma$-ray emission might even be placed upstream of BH.
But this is also unlikely, as the presence of a free-free absorbing torus \citep[see, for example][]{Walker00, Wajima20}, combined with the effect of Doppler boosting away from the line of sight, would not leave any $\gamma$-rays to be detected.
Therefore, based on the analysis presented here, we conclude that the source of the $\gamma$-rays is more likely to be located in the pc-scale jet, downstream of the core region of \C.

Attempts have already been made to interpret the short term $\gamma$-ray emission if it is not originating in the core region.
\cite{Hodgson18} proposed, for example, a turbulent, extreme multi-zone model \citep{Marscher14}.
A `split-flare-dissipate' scenario, where the splitting of components, in the pc-scale jet region due to magnetic reconnection \citep{Giannios13} causes `mini-jets' could also offer a potential explanation \citep{Hodgson21}.
Another possible explanation could be that a jet feature, the onset of which corresponded with the ejection of $\gamma$-rays upstream \citep[as described, for example, in the corresponding scenario put forth, by][]{LeonTavares11}, propagates further downstream and interacts with the shocked region / hot-spot C3 \citep[such an interaction between the jet and the circumnuclear environment in the C3 region was observed by][]{Kino18}, producing even more $\gamma$-rays.
Such a scenario would potentially explain both the leading and trailing emission of $\gamma$-rays with regard to the radio emission.

\subsection{Radio feature ejection \& $\gamma$-ray emission}\label{ssec:radio-gamma}

The association between emissions of radio features from the core region and total intensity radio flux increases is theoretically understood \citep{Marscher85} and has been observationally studied in the pc-scale jets of AGNs \citep{Savolainen02}.
In the work by \cite{Hodgson21}, a potential association between jet kinematics and $\gamma$-ray activity in investigated.
The authors found that the major $\gamma$-ray flares seem to be associated with the merging and splitting of jet components (at 43\,GHz) in the C3 region.

\cite{Paraschos22} found three core-ejected features (cross-identified at 43 and 86\,GHz), denoted as `F3', `F4', and `F5', which appear associated with outbursts in the centimetre and millimetre flux.
Specifically, F3, and F4 seem to correspond to the onset of the radio flares in $2014.7\pm0.5$ and $2016.7\pm0.5$, whereas F5 with the decay of the latter flare.
Based on the DCF analysis of the $\gamma$-rays and the millimetre radio flux, the time lags between them are $\tau_1=-1.6\pm0.4$ years and $\tau_2=1.5\pm0.5$ years.
If the $\gamma$-ray flux trails the radio flux, as suggested in Sect.~\ref{ssec:gamma}, the time lag between them is positive and corresponds to $\tau_2$.
Based on this assumption and their ejection times, presented in \cite{Paraschos22}, features F3, F4, and F5 would be expected to cause $\gamma$-ray flares in $2013.3\pm0.7$, $2017.4\pm0.7$, and $2019.4\pm0.7$ respectively.
Following the naming convention in \cite{Hodgson18} we also identified the flare called `G1'\footnote{The flux peak G1 consists of two peaks in the GPR curve, with a time delay of $\sim0.3$\,yrs in-between. We averaged them to determine the flux peak.}, with a point of ejection in $2013.9\pm0.1$, which matches, within the error budget, the ejection time of a $\gamma$-ray flare associated with F3.
Furthermore, the highest intensity flare of the $\gamma$-rays, denoted as `G5' in Fig.~\ref{fig:DCF_gamma}, was ejected in $2018.3\pm0.1$, pointing to an association with F4.
F5, however, does not seem to be clearly associated with a $\gamma$-ray flare.
It should be noted here, that besides the major flares G1, `G2', `G3', `G4' \citep[][see also middle panel in Fig.~\ref{fig:DCF_gamma}]{Hodgson18} and `G0', G5, many smaller flares are also observed.

If, on the other hand, the radio flux trails the $\gamma$-ray flux, the expected time lag between them is negative and corresponds to $\tau_1$.
In the following we explore the time lags between the different flares.
The main flares in the millimetre radio flux are `Y0', `Y1', and `Y34' and they are denoted in the bottom panel of Fig.~\ref{fig:DCF_gamma}.
Using the GPR technique, we calculated that they peaked in $2012.2\pm0.1$, $2014.7\pm0.1$, and $2016.6\pm0.1$ respectively.
Accordingly, we computed the flare times of the $\gamma$-ray flares G0, G1, G3, and G4.
They peaked in $2010.6\pm0.1$, $2013.9\pm0.1$, $2015.1\pm0.1$, $2015.8\pm0.1$ respectively.
All flares discussed here are summarised in Table~\ref{table:Flares}.
The time lag between G0 and Y0 is $\sim1.6$\,yrs and this event might be associated with the ejection of component F3.
The time lag between G1 and Y1 is $\sim0.8$\,yrs.
G3 and G4 could both be associated with Y34, since they appeared temporally close to each other. 
The two time lags are $\sim1.5$\,yrs and $\sim0.8$\,yrs respectively.
This event might be associated with the ejection of component F4.
Our results hint at a possible association between jet features and $\gamma$-ray flares, although a conclusive relation is lacking.
Overall, in this scenario, we find a consistent time lag between the $\gamma$-ray and radio emission of the order of $\sim1.2$\,yrs for these three flares.
We note with interest that, if indeed the radio flux is trailing the $\gamma$-rays, G5 has no corresponding radio flare counterpart, perhaps indicating that \C\ is undergoing intrinsic changes in the inner pc-region, leading to this new behaviour.

\subsection{Intrinsic jet conditions}\label{ssec:lambda-sigma}

Core shift measurements can provide an insight into jet launching by means of the multiplicity parameter $\lambda$ and the Michel magnetisation $\sigma_\textrm{M}$ \citep{Nokhrina15}.
The former is defined as the ratio of the electron number density to the Goldreich-Julian number density \citep{Goldreich69}.
The latter is a measure of the magnetisation at the base of an outflow \citep{Michel69} and provides an upper limit to the bulk Lorentz factor $\Gamma$ of an MHD outflow.
Under the assumption of equipartition, \cite{Nokhrina15} calculated two equations, which connect $\lambda$ and $\sigma_\textrm{M}$ to the measured core shift \citep[see Eqs. 5 and 29 in][]{Nokhrina15}.
In the saturation regime \citep{Beskin10} and for a slightly magnetised flux (i.e. when the Poynting flux has been converted almost entirely into particle kinetic energy), if $\sigma_\textrm{M}\sim10-10^3$ and $\lambda\sim10^{10}-10^{13}$, then the accretion region is hot enough to facilitate the production of electron-positron pair plasma through two-photon collisions.
These photons originate from the inner part of the accretion disc \citep{Blandford77, Moscibrodzka11, Nokhrina15}.
On the other hand, when the Goldreich-Julian number density vanishes, and $\lambda\sim10-100$ and $\sigma_\textrm{M}\sim10^{10}-10^{12}$, electron-positron pair plasma can also be created in regions of the magnetosphere in the vicinity of a black hole, as shown by \cite{Hirotani98}.

As shown in Sect.~\ref{ssec:timelags}, the equipartition assumption holds and we can therefore calculate $\lambda$ and $\sigma_\textrm{M}$.
For the viewing angle, $\gamma_\textrm{max}$, $\gamma_\textrm{min}$, jet opening angle, jet power ($P_\textrm{jet}$) and optically thin spectral index gradient we used the value ranges listed in Table~\ref{table:Parameters}.
This results in a slightly different set of equations than Eqs.~31 and 32 in \cite{Nokhrina15}, since they assumed that $\alpha_\textrm{thin} = -0.5$.
For the ratio of the number density of the emitting particles to the MHD flow number density ($\xi$), we assumed a value of $\xi = 0.01$ \citep{Sironi13}, following \cite{Nokhrina15}.
Finally, for the core position offset we used the value ranges listed in Table~\ref{table:Output}.
To arrive at a final value range for $\lambda$ and $\sigma_\textrm{M}$, we averaged the measurements for all frequencies.
We find that $\lambda=\multpar$ and $\sigma=\michel$, which is consistent with the assumption of a jet launched by the \cite{Blandford77} mechanism .

\begin{threeparttable}[h]
\caption{Identified flares with respective peak times.}      
\label{table:Flares}
\centering    
\begin{tabular}{cc}
Identifier & Year \\
\hline\hline
 G0  &  2010.6\,$\pm$\,0.1\\
 G1  &  2013.9\,$\pm$\,0.1\\
 G3  &  2015.1\,$\pm$\,0.1\\
 G4  &  2010.8\,$\pm$\,0.1\\
 G5  &  2018.3\,$\pm$\,0.1\\
 Y0  &  2012.2\,$\pm$\,0.1\\
 Y1  &  2014.7\,$\pm$\,0.1\\
 Y34 &  2016.6\,$\pm$\,0.1\\
\hline
\end{tabular}
\end{threeparttable}

\section{Conclusions}\label{sec:Conclusions}

We have used total intensity centimetre and millimetre radio light curves to determine the core shift and magnetic field strength of \C.
Furthermore, we have used the $\gamma$-ray light curve to investigate the association between $\gamma$-ray bursts and radio component ejections and to explore the jet launching mechanism.
Our major findings can be summarised as follows.

\begin{enumerate}
    \item We estimate the jet particle and magnetic field energy densities to be in equipartition ($k_\textrm{r}=\kav$).
    \item Using the techniques of GPR and DCF, we constrained the jet apex to be $z_\textrm{91.5\,GHz}=\znf$.
    This value compares well to the results presented in \cite{Paraschos21} and \cite{Oh22}.
    \item At the jet apex, the magnetic field strength is $B_\textrm{core}^\textrm{91.5\,GHz} = \Bnoround$\,G.
    \item Our analysis shows that the $\gamma$-ray and radio emission are likely associated.
    \item Features F3 and F4 \citep{Paraschos22} might be connected to $\gamma$-ray flares; F5 does not seem to have a radio flux counterpart.
    \item We computed the multiplicity factor and Michel magnetisation to be $\lambda = \multpar$ and $\sigma_\textrm{M} = \michel$ respectively.
    These values are in agreement with the \cite{Blandford77} jet launching mechanism.
\end{enumerate}

\begin{acknowledgements}
      We thank the anonymous referee for the detailed comments, which improved this manuscript.
      G. F. Paraschos is supported for this research by the International Max-Planck Research School (IMPRS) for Astronomy and Astrophysics at the University of Bonn and Cologne. 
      J.-Y. Kim acknowledges support from the National Research Foundation (NRF) of Korea (grant no. 2022R1C1C1005255).
      This work makes use of 37 GHz, and 230 and 345\,GHz light curves kindly provided by the Aalto University Mets\"ahovi Radio Observatory and the Submillimeter Array (SMA), respectively.
      The SMA is a joint project between the Smithsonian Astrophysical Observatory and the Academia Sinica Institute of Astronomy and Astrophysics and is funded by the Smithsonian Institution and the Academia Sinica.
      Maunakea, the location of the SMA, is a culturally important site for the indigenous Hawaiian people; we are privileged to study the cosmos from its summit.
      This research has made use of data from the University of Michigan Radio Astronomy Observatory which has been supported by the University of Michigan and by a series of grants from the National Science Foundation, most recently AST-0607523.
      This work makes use of the Swinburne University of Technology software correlator, developed as part of the Australian Major National Research Facilities Programme and operated under licence. 
    This research has made use of the NASA/IPAC Extragalactic Database (NED), which is operated by the Jet Propulsion Laboratory, California Institute of Technology, under contract with the National Aeronautics and Space Administration. This research has also made use of NASA's Astrophysics Data System Bibliographic Services. 
      This research has also made use of data from the OVRO 40-m monitoring program \citep{Richards11}, supported by private funding from the California Institute of Technology and the Max Planck Institute for Radio Astronomy, and by NASA grants NNX08AW31G, NNX11A043G, and NNX14AQ89G and NSF grants AST-0808050 and AST-1109911.
      S.K. acknowledges support from the European Research Council (ERC) under the European Unions Horizon 2020 research and innovation programme under grant agreement No.~771282.
      Finally, this research made use of the following python packages: {\it numpy} \citep{Harris20}, {\it scipy} \citep{2020SciPy-NMeth}, {\it matplotlib} \citep{Hunter07}, {\it astropy} \citep{2013A&A...558A..33A, 2018AJ....156..123A} and {\it Uncertainties: a Python package for calculations with uncertainties}.
\end{acknowledgements}

\bibliographystyle{aa} 
\bibliography{sources}

\begin{thebibliography}{88}
\expandafter\ifx\csname natexlab\endcsname\relax\def\natexlab#1{#1}\fi

\bibitem[{{Abdo} {et~al.}(2009){Abdo}, {Ackermann}, {Ajello}, {Asano},
  {Baldini}, {Ballet}, {Barbiellini}, {Bastieri}, {Baughman}, {Bechtol},
  {Bellazzini}, {Blandford}, {Bloom}, {Bonamente}, {Borgland}, {Bregeon},
  {Brez}, {Brigida}, {Bruel}, {Burnett}, {Caliandro}, {Cameron}, {Caraveo},
  {Casandjian}, {Cavazzuti}, {Cecchi}, {Celotti}, {Chekhtman}, {Cheung},
  {Chiang}, {Ciprini}, {Claus}, {Cohen-Tanugi}, {Colafrancesco}, {Cominsky},
  {Conrad}, {Costamante}, {Dermer}, {de Angelis}, {de Palma}, {Digel},
  {Donato}, {do Couto e Silva}, {Drell}, {Dubois}, {Dumora}, {Farnier},
  {Favuzzi}, {Finke}, {Focke}, {Frailis}, {Fukazawa}, {Funk}, {Fusco},
  {Gargano}, {Georganopoulos}, {Germani}, {Giebels}, {Giglietto}, {Giordano},
  {Glanzman}, {Grenier}, {Grondin}, {Grove}, {Guillemot}, {Guiriec},
  {Hanabata}, {Harding}, {Hartman}, {Hayashida}, {Hays}, {Hughes},
  {J{\'o}hannesson}, {Johnson}, {Johnson}, {Johnson}, {Kadler}, {Kamae},
  {Kanai}, {Katagiri}, {Kataoka}, {Kawai}, {Kerr}, {Kn{\"o}dlseder}, {Kuehn},
  {Kuss}, {Latronico}, {Lemoine-Goumard}, {Longo}, {Loparco}, {Lott},
  {Lovellette}, {Lubrano}, {Madejski}, {Makeev}, {Mazziotta}, {McEnery},
  {Meurer}, {Michelson}, {Mitthumsiri}, {Mizuno}, {Moiseev}, {Monte},
  {Monzani}, {Morselli}, {Moskalenko}, {Murgia}, {Nakamori}, {Nolan}, {Norris},
  {Nuss}, {Ohsugi}, {Omodei}, {Orlando}, {Ormes}, {Paneque}, {Panetta},
  {Parent}, {Pepe}, {Pesce-Rollins}, {Piron}, {Porter}, {Rain{\`o}}, {Razzano},
  {Reimer}, {Reimer}, {Reposeur}, {Ritz}, {Rodriguez}, {Romani}, {Ryde},
  {Sadrozinski}, {Sambruna}, {Sanchez}, {Sander}, {Sato}, {Parkinson},
  {Sgr{\`o}}, {Smith}, {Smith}, {Spandre}, {Spinelli}, {Starck}, {Strickman},
  {Strong}, {Suson}, {Tajima}, {Takahashi}, {Takahashi}, {Tanaka}, {Taylor},
  {Thayer}, {Thompson}, {Torres}, {Tosti}, {Uchiyama}, {Usher}, {Vilchez},
  {Vitale}, {Waite}, {Wood}, {Ylinen}, {Ziegler}, {Aller}, {Aller},
  {Kellermann}, {Kovalev}, {Kovalev}, {Lister}, \& {Pushkarev}}]{Abdo09}
{Abdo}, A.~A., {Ackermann}, M., {Ajello}, M., {et~al.} 2009, \apj, 699, 31

\bibitem[{{Ambikasaran} {et~al.}(2015){Ambikasaran}, {Foreman-Mackey},
  {Greengard}, {Hogg}, \& {O'Neil}}]{Ambikasaran15}
{Ambikasaran}, S., {Foreman-Mackey}, D., {Greengard}, L., {Hogg}, D.~W., \&
  {O'Neil}, M. 2015, IEEE Transactions on Pattern Analysis and Machine
  Intelligence, 38, 252

\bibitem[{{Astropy Collaboration} {et~al.}(2018){Astropy Collaboration},
  {Price-Whelan}, {Sip{\H{o}}cz}, {G{\"u}nther}, {Lim}, {Crawford}, {Conseil},
  {Shupe}, {Craig}, {Dencheva}, {Ginsburg}, {VanderPlas}, {Bradley},
  {P{\'e}rez-Su{\'a}rez}, {de Val-Borro}, {Aldcroft}, {Cruz}, {Robitaille},
  {Tollerud}, {Ardelean}, {Babej}, {Bach}, {Bachetti}, {Bakanov}, {Bamford},
  {Barentsen}, {Barmby}, {Baumbach}, {Berry}, {Biscani}, {Boquien}, {Bostroem},
  {Bouma}, {Brammer}, {Bray}, {Breytenbach}, {Buddelmeijer}, {Burke},
  {Calderone}, {Cano Rodr{\'\i}guez}, {Cara}, {Cardoso}, {Cheedella}, {Copin},
  {Corrales}, {Crichton}, {D'Avella}, {Deil}, {Depagne}, {Dietrich}, {Donath},
  {Droettboom}, {Earl}, {Erben}, {Fabbro}, {Ferreira}, {Finethy}, {Fox},
  {Garrison}, {Gibbons}, {Goldstein}, {Gommers}, {Greco}, {Greenfield},
  {Groener}, {Grollier}, {Hagen}, {Hirst}, {Homeier}, {Horton}, {Hosseinzadeh},
  {Hu}, {Hunkeler}, {Ivezi{\'c}}, {Jain}, {Jenness}, {Kanarek}, {Kendrew},
  {Kern}, {Kerzendorf}, {Khvalko}, {King}, {Kirkby}, {Kulkarni}, {Kumar},
  {Lee}, {Lenz}, {Littlefair}, {Ma}, {Macleod}, {Mastropietro}, {McCully},
  {Montagnac}, {Morris}, {Mueller}, {Mumford}, {Muna}, {Murphy}, {Nelson},
  {Nguyen}, {Ninan}, {N{\"o}the}, {Ogaz}, {Oh}, {Parejko}, {Parley}, {Pascual},
  {Patil}, {Patil}, {Plunkett}, {Prochaska}, {Rastogi}, {Reddy Janga},
  {Sabater}, {Sakurikar}, {Seifert}, {Sherbert}, {Sherwood-Taylor}, {Shih},
  {Sick}, {Silbiger}, {Singanamalla}, {Singer}, {Sladen}, {Sooley},
  {Sornarajah}, {Streicher}, {Teuben}, {Thomas}, {Tremblay}, {Turner},
  {Terr{\'o}n}, {van Kerkwijk}, {de la Vega}, {Watkins}, {Weaver}, {Whitmore},
  {Woillez}, {Zabalza}, \& {Astropy Contributors}}]{2018AJ....156..123A}
{Astropy Collaboration}, {Price-Whelan}, A.~M., {Sip{\H{o}}cz}, B.~M., {et~al.}
  2018, \aj, 156, 123

\bibitem[{{Astropy Collaboration} {et~al.}(2013){Astropy Collaboration},
  {Robitaille}, {Tollerud}, {Greenfield}, {Droettboom}, {Bray}, {Aldcroft},
  {Davis}, {Ginsburg}, {Price-Whelan}, {Kerzendorf}, {Conley}, {Crighton},
  {Barbary}, {Muna}, {Ferguson}, {Grollier}, {Parikh}, {Nair}, {Unther},
  {Deil}, {Woillez}, {Conseil}, {Kramer}, {Turner}, {Singer}, {Fox}, {Weaver},
  {Zabalza}, {Edwards}, {Azalee Bostroem}, {Burke}, {Casey}, {Crawford},
  {Dencheva}, {Ely}, {Jenness}, {Labrie}, {Lim}, {Pierfederici}, {Pontzen},
  {Ptak}, {Refsdal}, {Servillat}, \& {Streicher}}]{2013A&A...558A..33A}
{Astropy Collaboration}, {Robitaille}, T.~P., {Tollerud}, E.~J., {et~al.} 2013,
  \aap, 558, A33

\bibitem[{{Atwood} {et~al.}(2009){Atwood}, {Abdo}, {Ackermann}, {Althouse},
  {Anderson}, {Axelsson}, {Baldini}, {Ballet}, {Band}, {Barbiellini},
  {Bartelt}, {Bastieri}, {Baughman}, {Bechtol}, {B{\'e}d{\'e}r{\`e}de},
  {Bellardi}, {Bellazzini}, {Berenji}, {Bignami}, {Bisello}, {Bissaldi},
  {Blandford}, {Bloom}, {Bogart}, {Bonamente}, {Bonnell}, {Borgland},
  {Bouvier}, {Bregeon}, {Brez}, {Brigida}, {Bruel}, {Burnett}, {Busetto},
  {Caliandro}, {Cameron}, {Caraveo}, {Carius}, {Carlson}, {Casandjian},
  {Cavazzuti}, {Ceccanti}, {Cecchi}, {Charles}, {Chekhtman}, {Cheung},
  {Chiang}, {Chipaux}, {Cillis}, {Ciprini}, {Claus}, {Cohen-Tanugi},
  {Condamoor}, {Conrad}, {Corbet}, {Corucci}, {Costamante}, {Cutini}, {Davis},
  {Decotigny}, {DeKlotz}, {Dermer}, {de Angelis}, {Digel}, {do Couto e Silva},
  {Drell}, {Dubois}, {Dumora}, {Edmonds}, {Fabiani}, {Farnier}, {Favuzzi},
  {Flath}, {Fleury}, {Focke}, {Funk}, {Fusco}, {Gargano}, {Gasparrini},
  {Gehrels}, {Gentit}, {Germani}, {Giebels}, {Giglietto}, {Giommi}, {Giordano},
  {Glanzman}, {Godfrey}, {Grenier}, {Grondin}, {Grove}, {Guillemot}, {Guiriec},
  {Haller}, {Harding}, {Hart}, {Hays}, {Healey}, {Hirayama}, {Hjalmarsdotter},
  {Horn}, {Hughes}, {J{\'o}hannesson}, {Johansson}, {Johnson}, {Johnson},
  {Johnson}, {Johnson}, {Kamae}, {Katagiri}, {Kataoka}, {Kavelaars}, {Kawai},
  {Kelly}, {Kerr}, {Klamra}, {Kn{\"o}dlseder}, {Kocian}, {Komin}, {Kuehn},
  {Kuss}, {Landriu}, {Latronico}, {Lee}, {Lee}, {Lemoine-Goumard}, {Lionetto},
  {Longo}, {Loparco}, {Lott}, {Lovellette}, {Lubrano}, {Madejski}, {Makeev},
  {Marangelli}, {Massai}, {Mazziotta}, {McEnery}, {Menon}, {Meurer},
  {Michelson}, {Minuti}, {Mirizzi}, {Mitthumsiri}, {Mizuno}, {Moiseev},
  {Monte}, {Monzani}, {Moretti}, {Morselli}, {Moskalenko}, {Murgia},
  {Nakamori}, {Nishino}, {Nolan}, {Norris}, {Nuss}, {Ohno}, {Ohsugi}, {Omodei},
  {Orlando}, {Ormes}, {Paccagnella}, {Paneque}, {Panetta}, {Parent}, {Pearce},
  {Pepe}, {Perazzo}, {Pesce-Rollins}, {Picozza}, {Pieri}, {Pinchera}, {Piron},
  {Porter}, {Poupard}, {Rain{\`o}}, {Rando}, {Rapposelli}, {Razzano}, {Reimer},
  {Reimer}, {Reposeur}, {Reyes}, {Ritz}, {Rochester}, {Rodriguez}, {Romani},
  {Roth}, {Russell}, {Ryde}, {Sabatini}, {Sadrozinski}, {Sanchez}, {Sander},
  {Sapozhnikov}, {Parkinson}, {Scargle}, {Schalk}, {Scolieri}, {Sgr{\`o}},
  {Share}, {Shaw}, {Shimokawabe}, {Shrader}, {Sierpowska-Bartosik}, {Siskind},
  {Smith}, {Smith}, {Spandre}, {Spinelli}, {Starck}, {Stephens}, {Strickman},
  {Strong}, {Suson}, {Tajima}, {Takahashi}, {Takahashi}, {Tanaka}, {Tenze},
  {Tether}, {Thayer}, {Thayer}, {Thompson}, {Tibaldo}, {Tibolla}, {Torres},
  {Tosti}, {Tramacere}, {Turri}, {Usher}, {Vilchez}, {Vitale}, {Wang},
  {Watters}, {Winer}, {Wood}, {Ylinen}, \& {Ziegler}}]{Atwood09}
{Atwood}, W.~B., {Abdo}, A.~A., {Ackermann}, M., {et~al.} 2009, \apj, 697, 1071

\bibitem[{{Bach} {et~al.}(2006){Bach}, {Villata}, {Raiteri}, {Agudo}, {Aller},
  {Aller}, {Denn}, {G{\'o}mez}, {Jorstad}, {Marscher}, {Mutel}, \&
  {Ter{\"a}sranta}}]{Bach06}
{Bach}, U., {Villata}, M., {Raiteri}, C.~M., {et~al.} 2006, \aap, 456, 105

\bibitem[{{Beskin}(2010)}]{Beskin10}
{Beskin}, V.~S. 2010, Physics Uspekhi, 53, 1199

\bibitem[{{Blandford} \& {K{\"o}nigl}(1979)}]{Blandford79}
{Blandford}, R.~D. \& {K{\"o}nigl}, A. 1979, \apj, 232, 34

\bibitem[{{Blandford} \& {Znajek}(1977)}]{Blandford77}
{Blandford}, R.~D. \& {Znajek}, R.~L. 1977, \mnras, 179, 433

\bibitem[{{Boccardi} {et~al.}(2016){Boccardi}, {Krichbaum}, {Bach}, {Mertens},
  {Ros}, {Alef}, \& {Zensus}}]{Boccardi16}
{Boccardi}, B., {Krichbaum}, T.~P., {Bach}, U., {et~al.} 2016, \aap, 585, A33

\bibitem[{{Britzen} {et~al.}(2019){Britzen}, {Fendt}, {Zaja{\v{c}}ek}, {Jaron},
  {Pashchenko}, {Aller}, \& {Aller}}]{Britzen19}
{Britzen}, S., {Fendt}, C., {Zaja{\v{c}}ek}, M., {et~al.} 2019, Galaxies, 7, 72

\bibitem[{{Chidiac} {et~al.}(2016){Chidiac}, {Rani}, {Krichbaum}, {Angelakis},
  {Fuhrmann}, {Nestoras}, {Zensus}, {Sievers}, {Ungerechts}, {Itoh},
  {Fukazawa}, {Uemura}, {Sasada}, {Gurwell}, \& {Fedorova}}]{Chidiac16}
{Chidiac}, C., {Rani}, B., {Krichbaum}, T.~P., {et~al.} 2016, \aap, 590, A61

\bibitem[{{Croke} \& {Gabuzda}(2008)}]{Croke08}
{Croke}, S.~M. \& {Gabuzda}, D.~C. 2008, \mnras, 386, 619

\bibitem[{{Dutson} {et~al.}(2014){Dutson}, {Edge}, {Hinton}, {Hogan},
  {Gurwell}, \& {Alston}}]{Dutson14}
{Dutson}, K.~L., {Edge}, A.~C., {Hinton}, J.~A., {et~al.} 2014, \mnras, 442,
  2048

\bibitem[{{Edelson} \& {Krolik}(1988)}]{Edelson88}
{Edelson}, R.~A. \& {Krolik}, J.~H. 1988, \apj, 333, 646

\bibitem[{{Fermi Large Area Telescope Collaboration}(2021)}]{Kocevski21}
{Fermi Large Area Telescope Collaboration}. 2021, The Astronomer's Telegram,
  15110, 1

\bibitem[{{Foreman-Mackey} {et~al.}(2013){Foreman-Mackey}, {Conley},
  {Meierjurgen Farr}, {Hogg}, {Lang}, {Marshall}, {Price-Whelan}, {Sanders}, \&
  {Zuntz}}]{Foreman13}
{Foreman-Mackey}, D., {Conley}, A., {Meierjurgen Farr}, W., {et~al.} 2013,
  {emcee: The MCMC Hammer}, Astrophysics Source Code Library, record
  ascl:1303.002

\bibitem[{Foreman-Mackey {et~al.}(2021)Foreman-Mackey, Price-Whelan, Vousden,
  Ryan, Pitkin, Zabalza, jsheyl, Smith, Ashton, Singer, Smith, Rice, Nitz,
  Brewer, Sipőcz, Hogg, Gentry, Rein, Madan, Czekala, Tocknell, Barbary,
  Prechelt, Hoyer, Caswell, Kerzendorf, \& Cruz}]{Foreman21}
Foreman-Mackey, D., Price-Whelan, A., Vousden, W., {et~al.} 2021,
  dfm/corner.py: corner.py v.2.2.1

\bibitem[{{Fromm} {et~al.}(2011){Fromm}, {Perucho}, {Ros}, {Savolainen},
  {Lobanov}, {Zensus}, {Aller}, {Aller}, {Gurwell}, \&
  {L{\"a}hteenm{\"a}ki}}]{Fromm11}
{Fromm}, C.~M., {Perucho}, M., {Ros}, E., {et~al.} 2011, \aap, 531, A95

\bibitem[{{Fromm} {et~al.}(2013){Fromm}, {Ros}, {Perucho}, {Savolainen},
  {Mimica}, {Kadler}, {Lobanov}, \& {Zensus}}]{Fromm13}
{Fromm}, C.~M., {Ros}, E., {Perucho}, M., {et~al.} 2013, \aap, 557, A105

\bibitem[{{Fuhrmann} {et~al.}(2014){Fuhrmann}, {Larsson}, {Chiang},
  {Angelakis}, {Zensus}, {Nestoras}, {Krichbaum}, {Ungerechts}, {Sievers},
  {Pavlidou}, {Readhead}, {Max-Moerbeck}, \& {Pearson}}]{Fuhrmann14}
{Fuhrmann}, L., {Larsson}, S., {Chiang}, J., {et~al.} 2014, \mnras, 441, 1899

\bibitem[{{Ghisellini} {et~al.}(2005){Ghisellini}, {Tavecchio}, \&
  {Chiaberge}}]{Ghisellini05}
{Ghisellini}, G., {Tavecchio}, F., \& {Chiaberge}, M. 2005, \aap, 432, 401

\bibitem[{{Giannios}(2013)}]{Giannios13}
{Giannios}, D. 2013, \mnras, 431, 355

\bibitem[{{Giovannini} {et~al.}(2018){Giovannini}, {Savolainen}, {Orienti},
  {Nakamura}, {Nagai}, {Kino}, {Giroletti}, {Hada}, {Bruni}, {Kovalev},
  {Anderson}, {D'Ammando}, {Hodgson}, {Honma}, {Krichbaum}, {Lee}, {Lico},
  {Lisakov}, {Lobanov}, {Petrov}, {Sohn}, {Sokolovsky}, {Voitsik}, {Zensus}, \&
  {Tingay}}]{Giovannini18}
{Giovannini}, G., {Savolainen}, T., {Orienti}, M., {et~al.} 2018, Nature
  Astronomy, 2, 472

\bibitem[{{Goldreich} \& {Julian}(1969)}]{Goldreich69}
{Goldreich}, P. \& {Julian}, W.~H. 1969, \apj, 157, 869

\bibitem[{{Guirado} {et~al.}(1995){Guirado}, {Marcaide}, {Alberdi}, {Elosegui},
  {Ratner}, {Shapiro}, {Kilger}, {Mantovani}, {Venturi}, {Rius}, {Ros},
  {Trigilio}, \& {Whitney}}]{Guirado95}
{Guirado}, J.~C., {Marcaide}, J.~M., {Alberdi}, A., {et~al.} 1995, \aj, 110,
  2586

\bibitem[{Harris {et~al.}(2020)Harris, Millman, van~der Walt, Gommers,
  Virtanen, Cournapeau, Wieser, Taylor, Berg, Smith, Kern, Picus, Hoyer, van
  Kerkwijk, Brett, Haldane, del R{'{\i}}o, Wiebe, Peterson,
  G{'{e}}rard-Marchant, Sheppard, Reddy, Weckesser, Abbasi, Gohlke, \&
  Oliphant}]{Harris20}
Harris, C.~R., Millman, K.~J., van~der Walt, S.~J., {et~al.} 2020, Nature, 585,
  357

\bibitem[{{Hirotani}(2005)}]{Hirotani05}
{Hirotani}, K. 2005, \apj, 619, 73

\bibitem[{{Hirotani} \& {Okamoto}(1998)}]{Hirotani98}
{Hirotani}, K. \& {Okamoto}, I. 1998, \apj, 497, 563

\bibitem[{{Hodgson} {et~al.}(2018){Hodgson}, {Rani}, {Lee}, {Algaba}, {Kino},
  {Trippe}, {Park}, {Zhao}, {Byun}, {Kang}, {Kim}, {Kim}, {Kim}, {Miyazaki},
  {Wajima}, {Oh}, {Kim}, \& {Gurwell}}]{Hodgson18}
{Hodgson}, J.~A., {Rani}, B., {Lee}, S.-S., {et~al.} 2018, \mnras, 475, 368

\bibitem[{{Hodgson} {et~al.}(2021){Hodgson}, {Rani}, {Oh}, {Marscher},
  {Jorstad}, {Mizuno}, {Park}, {Lee}, {Trippe}, \& {Mertens}}]{Hodgson21}
{Hodgson}, J.~A., {Rani}, B., {Oh}, J., {et~al.} 2021, \apj, 914, 43

\bibitem[{{Hovatta} {et~al.}(2007){Hovatta}, {Tornikoski}, {Lainela}, {Lehto},
  {Valtaoja}, {Torniainen}, {Aller}, \& {Aller}}]{Hovatta07}
{Hovatta}, T., {Tornikoski}, M., {Lainela}, M., {et~al.} 2007, \aap, 469, 899

\bibitem[{Hunter(2007)}]{Hunter07}
Hunter, J.~D. 2007, Computing in Science \& Engineering, 9, 90

\bibitem[{{Jorstad} {et~al.}(2001){Jorstad}, {Marscher}, {Mattox}, {Aller},
  {Aller}, {Wehrle}, \& {Bloom}}]{Jorstad01}
{Jorstad}, S.~G., {Marscher}, A.~P., {Mattox}, J.~R., {et~al.} 2001, \apj, 556,
  738

\bibitem[{{Karamanavis} {et~al.}(2016){Karamanavis}, {Fuhrmann}, {Krichbaum},
  {Angelakis}, {Hodgson}, {Nestoras}, {Myserlis}, {Zensus}, {Sievers}, \&
  {Ciprini}}]{Karamanavis16}
{Karamanavis}, V., {Fuhrmann}, L., {Krichbaum}, T.~P., {et~al.} 2016, \aap,
  586, A60

\bibitem[{{Kim} {et~al.}(2019){Kim}, {Krichbaum}, {Marscher}, {Jorstad},
  {Agudo}, {Thum}, {Hodgson}, {MacDonald}, {Ros}, {Lu}, {Bremer}, {de Vicente},
  {Lindqvist}, {Trippe}, \& {Zensus}}]{Kim19}
{Kim}, J.~Y., {Krichbaum}, T.~P., {Marscher}, A.~P., {et~al.} 2019, \aap, 622,
  A196

\bibitem[{{Kino} {et~al.}(2018){Kino}, {Wajima}, {Kawakatu}, {Nagai},
  {Orienti}, {Giovannini}, {Hada}, {Niinuma}, \& {Giroletti}}]{Kino18}
{Kino}, M., {Wajima}, K., {Kawakatu}, N., {et~al.} 2018, \apj, 864, 118

\bibitem[{{K\"onigl}(1981)}]{Konigl81}
{K\"onigl}, A. 1981, \apj, 243, 700

\bibitem[{{Kovalev} {et~al.}(2008){Kovalev}, {Lobanov}, {Pushkarev}, \&
  {Zensus}}]{Kovalev08}
{Kovalev}, Y.~Y., {Lobanov}, A.~P., {Pushkarev}, A.~B., \& {Zensus}, J.~A.
  2008, \aap, 483, 759

\bibitem[{{Kramarenko} {et~al.}(2022){Kramarenko}, {Pushkarev}, {Kovalev},
  {Lister}, {Hovatta}, \& {Savolainen}}]{Kramarenko22}
{Kramarenko}, I.~G., {Pushkarev}, A.~B., {Kovalev}, Y.~Y., {et~al.} 2022,
  \mnras, 510, 469

\bibitem[{{Kramer} \& {MacDonald}(2021)}]{Kramer21}
{Kramer}, J.~A. \& {MacDonald}, N.~R. 2021, \aap, 656, A143

\bibitem[{{Kudryavtseva} {et~al.}(2011){Kudryavtseva}, {Gabuzda}, {Aller}, \&
  {Aller}}]{Kudryavtseva11}
{Kudryavtseva}, N.~A., {Gabuzda}, D.~C., {Aller}, M.~F., \& {Aller}, H.~D.
  2011, \mnras, 415, 1631

\bibitem[{{Kutkin} {et~al.}(2018){Kutkin}, {Pashchenko}, {Lisakov}, {Voytsik},
  {Sokolovsky}, {Kovalev}, {Lobanov}, {Ipatov}, {Aller}, {Aller},
  {Lahteenmaki}, {Tornikoski}, \& {Gurvits}}]{Kutkin18}
{Kutkin}, A.~M., {Pashchenko}, I.~N., {Lisakov}, M.~M., {et~al.} 2018, \mnras,
  475, 4994

\bibitem[{{Kutkin} {et~al.}(2014){Kutkin}, {Sokolovsky}, {Lisakov}, {Kovalev},
  {Savolainen}, {Voytsik}, {Lobanov}, {Aller}, {Aller}, {Lahteenmaki},
  {Tornikoski}, {Volvach}, \& {Volvach}}]{Kutkin14}
{Kutkin}, A.~M., {Sokolovsky}, K.~V., {Lisakov}, M.~M., {et~al.} 2014, \mnras,
  437, 3396

\bibitem[{{L{\"a}hteenm{\"a}ki} \& {Valtaoja}(2003)}]{Lahteenmaki03}
{L{\"a}hteenm{\"a}ki}, A. \& {Valtaoja}, E. 2003, \apj, 590, 95

\bibitem[{{Le{\'o}n-Tavares} {et~al.}(2011){Le{\'o}n-Tavares}, {Valtaoja},
  {Tornikoski}, {L{\"a}hteenm{\"a}ki}, \& {Nieppola}}]{LeonTavares11}
{Le{\'o}n-Tavares}, J., {Valtaoja}, E., {Tornikoski}, M.,
  {L{\"a}hteenm{\"a}ki}, A., \& {Nieppola}, E. 2011, \aap, 532, A146

\bibitem[{{Linhoff} {et~al.}(2021){Linhoff}, {Sandrock}, {Kadler},
  {Els{\"a}sser}, \& {Rhode}}]{Linhoff21}
{Linhoff}, L., {Sandrock}, A., {Kadler}, M., {Els{\"a}sser}, D., \& {Rhode}, W.
  2021, \mnras, 500, 4671

\bibitem[{{Lobanov}(1998{\natexlab{a}})}]{Lobanov98a}
{Lobanov}, A.~P. 1998{\natexlab{a}}, \aaps, 132, 261

\bibitem[{{Lobanov}(1998{\natexlab{b}})}]{Lobanov98b}
{Lobanov}, A.~P. 1998{\natexlab{b}}, \aap, 330, 79

\bibitem[{{MAGIC Collaboration} {et~al.}(2018){MAGIC Collaboration}, {Ansoldi},
  {Antonelli}, {Arcaro}, {Baack}, {Babi{\'c}}, {Banerjee}, {Bangale}, {Barres
  de Almeida}, {Barrio}, {Becerra Gonz{\'a}lez}, {Bednarek}, {Bernardini},
  {Berse}, {Berti}, {Bhattacharyya}, {Bigongiari}, {Biland}, {Blanch},
  {Bonnoli}, {Carosi}, {Ceribella}, {Chatterjee}, {Colak}, {Colin}, {Colombo},
  {Contreras}, {Cortina}, {Covino}, {Cumani}, {D'Elia}, {da Vela}, {Dazzi}, {de
  Angelis}, {de Lotto}, {Delfino}, {Delgado}, {di Pierro}, {Dom{\'\i}nguez},
  {Dominis Prester}, {Dorner}, {Doro}, {Einecke}, {Elsaesser}, {Fallah
  Ramazani}, {Fattorini}, {Fern{\'a}ndez-Barral}, {Ferrara}, {Fidalgo},
  {Foffano}, {Fonseca}, {Font}, {Fruck}, {Galindo}, {Gallozzi}, {Garc{\'\i}a
  L{\'o}pez}, {Garczarczyk}, {Gaug}, {Giammaria}, {Godinovi{\'c}}, {Gora},
  {Guberman}, {Hadasch}, {Hahn}, {Hassan}, {Hayashida}, {Herrera}, {Hoang},
  {Hose}, {Hrupec}, {Ishio}, {Konno}, {Kubo}, {Kushida}, {Lamastra}, {Lelas},
  {Leone}, {Lindfors}, {Lombardi}, {Longo}, {L{\'o}pez}, {Maggio}, {Majumdar},
  {Makariev}, {Maneva}, {Manganaro}, {Mannheim}, {Maraschi}, {Mariotti},
  {Mart{\'\i}nez}, {Masuda}, {Mazin}, {Mielke}, {Minev}, {Miranda}, {Mirzoyan},
  {Moralejo}, {Moreno}, {Moretti}, {Nagayoshi}, {Neustroev}, {Niedzwiecki},
  {Nievas Rosillo}, {Nigro}, {Nilsson}, {Ninci}, {Nishijima}, {Noda},
  {Nogu{\'e}s}, {Paiano}, {Palacio}, {Paneque}, {Paoletti}, {Paredes},
  {Pedaletti}, {Pe{\~n}il}, {Peresano}, {Persic}, {Pfrang}, {Prada Moroni},
  {Prandini}, {Puljak}, {Garcia}, {Reichardt}, {Rhode}, {Rib{\'o}}, {Rico},
  {Righi}, {Rugliancich}, {Saha}, {Saito}, {Satalecka}, {Schweizer}, {Sitarek},
  {{\v{S}}nidari{\'c}}, {Sobczynska}, {Stamerra}, {Strzys}, {Suri{\'c}},
  {Takahashi}, {Tavecchio}, {Temnikov}, {Terzi{\'c}}, {Teshima},
  {Torres-Alb{\`a}}, {Tsujimoto}, {Vanzo}, {Vazquez Acosta}, {Vovk}, {Ward},
  {Will}, {Zari{\'c}}, {Glawion}, {Takalo}, \& {Jormanainen}}]{Magic18}
{MAGIC Collaboration}, {Ansoldi}, S., {Antonelli}, L.~A., {et~al.} 2018, \aap,
  617, A91

\bibitem[{{Marcaide} \& {Shapiro}(1984)}]{Marcaide84}
{Marcaide}, J.~M. \& {Shapiro}, I.~I. 1984, \apj, 276, 56

\bibitem[{{Marscher}(2014)}]{Marscher14}
{Marscher}, A.~P. 2014, \apj, 780, 87

\bibitem[{{Marscher} \& {Gear}(1985)}]{Marscher85}
{Marscher}, A.~P. \& {Gear}, W.~K. 1985, \apj, 298, 114

\bibitem[{{Max-Moerbeck} {et~al.}(2014){Max-Moerbeck}, {Hovatta}, {Richards},
  {King}, {Pearson}, {Readhead}, {Reeves}, {Shepherd}, {Stevenson},
  {Angelakis}, {Fuhrmann}, {Grainge}, {Pavlidou}, {Romani}, \&
  {Zensus}}]{MaxMoerbeck14}
{Max-Moerbeck}, W., {Hovatta}, T., {Richards}, J.~L., {et~al.} 2014, \mnras,
  445, 428

\bibitem[{{Mertens} \& {Lobanov}(2015)}]{Mertens15}
{Mertens}, F. \& {Lobanov}, A. 2015, \aap, 574, A67

\bibitem[{{Mertens} {et~al.}(2018){Mertens}, {Ghosh}, \&
  {Koopmans}}]{Mertens18}
{Mertens}, F.~G., {Ghosh}, A., \& {Koopmans}, L.~V.~E. 2018, \mnras, 478, 3640

\bibitem[{{Michel}(1969)}]{Michel69}
{Michel}, F.~C. 1969, \apj, 158, 727

\bibitem[{{Mo{\'s}cibrodzka} {et~al.}(2011){Mo{\'s}cibrodzka}, {Gammie},
  {Dolence}, \& {Shiokawa}}]{Moscibrodzka11}
{Mo{\'s}cibrodzka}, M., {Gammie}, C.~F., {Dolence}, J.~C., \& {Shiokawa}, H.
  2011, \apj, 735, 9

\bibitem[{{Nagai} {et~al.}(2016){Nagai}, {Chida}, {Kino}, {Orienti},
  {D'Ammando}, {Giovannini}, \& {Hiura}}]{Nagai16}
{Nagai}, H., {Chida}, H., {Kino}, M., {et~al.} 2016, Astronomische Nachrichten,
  337, 69

\bibitem[{{Nagai} {et~al.}(2014){Nagai}, {Haga}, {Giovannini}, {Doi},
  {Orienti}, {D'Ammando}, {Kino}, {Nakamura}, {Asada}, {Hada}, \&
  {Giroletti}}]{Nagai14}
{Nagai}, H., {Haga}, T., {Giovannini}, G., {et~al.} 2014, \apj, 785, 53

\bibitem[{{Nesterov} {et~al.}(1995){Nesterov}, {Lyuty}, \&
  {Valtaoja}}]{Nesterov95}
{Nesterov}, N.~S., {Lyuty}, V.~M., \& {Valtaoja}, E. 1995, \aap, 296, 628

\bibitem[{{Nokhrina} {et~al.}(2015){Nokhrina}, {Beskin}, {Kovalev}, \&
  {Zheltoukhov}}]{Nokhrina15}
{Nokhrina}, E.~E., {Beskin}, V.~S., {Kovalev}, Y.~Y., \& {Zheltoukhov}, A.~A.
  2015, \mnras, 447, 2726

\bibitem[{{Oh} {et~al.}(2022){Oh}, {Hodgson}, {Trippe}, {Krichbaum}, {Kam},
  {Paraschos}, {Kim}, {Rani}, {Sohn}, {Lee}, {Lico}, {Liuzzo}, {Bremer}, \&
  {Zensus}}]{Oh22}
{Oh}, J., {Hodgson}, J.~A., {Trippe}, S., {et~al.} 2022, \mnras, 509, 1024

\bibitem[{{Paraschos} {et~al.}(2021){Paraschos}, {Kim}, {Krichbaum}, \&
  {Zensus}}]{Paraschos21}
{Paraschos}, G.~F., {Kim}, J.~Y., {Krichbaum}, T.~P., \& {Zensus}, J.~A. 2021,
  \aap, 650, L18

\bibitem[{{Paraschos} {et~al.}(2022){Paraschos}, {Krichbaum}, {Kim}, {Hodgson},
  {Oh}, {Ros}, {Zensus}, {Marscher}, {Jorstad}, {Gurwell},
  {L{\"a}hteenm{\"a}ki}, {Tornikoski}, {Kiehlmann}, \&
  {Readhead}}]{Paraschos22}
{Paraschos}, G.~F., {Krichbaum}, T.~P., {Kim}, J.~Y., {et~al.} 2022, \aap, 665,
  A1

\bibitem[{{Pashchenko} {et~al.}(2020){Pashchenko}, {Plavin}, {Kutkin}, \&
  {Kovalev}}]{Pashchenko20}
{Pashchenko}, I.~N., {Plavin}, A.~V., {Kutkin}, A.~M., \& {Kovalev}, Y.~Y.
  2020, \mnras, 499, 4515

\bibitem[{{Planck Collaboration} {et~al.}(2016){Planck Collaboration}, {Ade},
  {Aghanim}, {Arnaud}, {Ashdown}, {Aumont}, {Baccigalupi}, {Banday},
  {Barreiro}, {Bartlett}, {Bartolo}, {Battaner}, {Battye}, {Benabed},
  {Beno{\^\i}t}, {Benoit-L{\'e}vy}, {Bernard}, {Bersanelli}, {Bielewicz},
  {Bock}, {Bonaldi}, {Bonavera}, {Bond}, {Borrill}, {Bouchet}, {Boulanger},
  {Bucher}, {Burigana}, {Butler}, {Calabrese}, {Cardoso}, {Catalano},
  {Challinor}, {Chamballu}, {Chary}, {Chiang}, {Chluba}, {Christensen},
  {Church}, {Clements}, {Colombi}, {Colombo}, {Combet}, {Coulais}, {Crill},
  {Curto}, {Cuttaia}, {Danese}, {Davies}, {Davis}, {de Bernardis}, {de Rosa},
  {de Zotti}, {Delabrouille}, {D{\'e}sert}, {Di Valentino}, {Dickinson},
  {Diego}, {Dolag}, {Dole}, {Donzelli}, {Dor{\'e}}, {Douspis}, {Ducout},
  {Dunkley}, {Dupac}, {Efstathiou}, {Elsner}, {En{\ss}lin}, {Eriksen},
  {Farhang}, {Fergusson}, {Finelli}, {Forni}, {Frailis}, {Fraisse},
  {Franceschi}, {Frejsel}, {Galeotta}, {Galli}, {Ganga}, {Gauthier}, {Gerbino},
  {Ghosh}, {Giard}, {Giraud-H{\'e}raud}, {Giusarma}, {Gjerl{\o}w},
  {Gonz{\'a}lez-Nuevo}, {G{\'o}rski}, {Gratton}, {Gregorio}, {Gruppuso},
  {Gudmundsson}, {Hamann}, {Hansen}, {Hanson}, {Harrison}, {Helou},
  {Henrot-Versill{\'e}}, {Hern{\'a}ndez-Monteagudo}, {Herranz}, {Hildebrandt},
  {Hivon}, {Hobson}, {Holmes}, {Hornstrup}, {Hovest}, {Huang}, {Huffenberger},
  {Hurier}, {Jaffe}, {Jaffe}, {Jones}, {Juvela}, {Keih{\"a}nen}, {Keskitalo},
  {Kisner}, {Kneissl}, {Knoche}, {Knox}, {Kunz}, {Kurki-Suonio}, {Lagache},
  {L{\"a}hteenm{\"a}ki}, {Lamarre}, {Lasenby}, {Lattanzi}, {Lawrence}, {Leahy},
  {Leonardi}, {Lesgourgues}, {Levrier}, {Lewis}, {Liguori}, {Lilje},
  {Linden-V{\o}rnle}, {L{\'o}pez-Caniego}, {Lubin}, {Mac{\'\i}as-P{\'e}rez},
  {Maggio}, {Maino}, {Mandolesi}, {Mangilli}, {Marchini}, {Maris}, {Martin},
  {Martinelli}, {Mart{\'\i}nez-Gonz{\'a}lez}, {Masi}, {Matarrese}, {McGehee},
  {Meinhold}, {Melchiorri}, {Melin}, {Mendes}, {Mennella}, {Migliaccio},
  {Millea}, {Mitra}, {Miville-Desch{\^e}nes}, {Moneti}, {Montier}, {Morgante},
  {Mortlock}, {Moss}, {Munshi}, {Murphy}, {Naselsky}, {Nati}, {Natoli},
  {Netterfield}, {N{\o}rgaard-Nielsen}, {Noviello}, {Novikov}, {Novikov},
  {Oxborrow}, {Paci}, {Pagano}, {Pajot}, {Paladini}, {Paoletti}, {Partridge},
  {Pasian}, {Patanchon}, {Pearson}, {Perdereau}, {Perotto}, {Perrotta},
  {Pettorino}, {Piacentini}, {Piat}, {Pierpaoli}, {Pietrobon}, {Plaszczynski},
  {Pointecouteau}, {Polenta}, {Popa}, {Pratt}, {Pr{\'e}zeau}, {Prunet},
  {Puget}, {Rachen}, {Reach}, {Rebolo}, {Reinecke}, {Remazeilles}, {Renault},
  {Renzi}, {Ristorcelli}, {Rocha}, {Rosset}, {Rossetti}, {Roudier},
  {Rouill{\'e} d'Orfeuil}, {Rowan-Robinson}, {Rubi{\~n}o-Mart{\'\i}n},
  {Rusholme}, {Said}, {Salvatelli}, {Salvati}, {Sandri}, {Santos},
  {Savelainen}, {Savini}, {Scott}, {Seiffert}, {Serra}, {Shellard}, {Spencer},
  {Spinelli}, {Stolyarov}, {Stompor}, {Sudiwala}, {Sunyaev}, {Sutton},
  {Suur-Uski}, {Sygnet}, {Tauber}, {Terenzi}, {Toffolatti}, {Tomasi},
  {Tristram}, {Trombetti}, {Tucci}, {Tuovinen}, {T{\"u}rler}, {Umana},
  {Valenziano}, {Valiviita}, {Van Tent}, {Vielva}, {Villa}, {Wade}, {Wandelt},
  {Wehus}, {White}, {White}, {Wilkinson}, {Yvon}, {Zacchei}, \&
  {Zonca}}]{Planck16}
{Planck Collaboration}, {Ade}, P.~A.~R., {Aghanim}, N., {et~al.} 2016, \aap,
  594, A13

\bibitem[{{Pushkarev} {et~al.}(2019){Pushkarev}, {Butuzova}, {Kovalev}, \&
  {Hovatta}}]{Pushkarev19}
{Pushkarev}, A.~B., {Butuzova}, M.~S., {Kovalev}, Y.~Y., \& {Hovatta}, T. 2019,
  \mnras, 482, 2336

\bibitem[{{Pushkarev} {et~al.}(2010){Pushkarev}, {Kovalev}, \&
  {Lister}}]{Pushkarev10}
{Pushkarev}, A.~B., {Kovalev}, Y.~Y., \& {Lister}, M.~L. 2010, \apjl, 722, L7

\bibitem[{{Ramakrishnan} {et~al.}(2015){Ramakrishnan}, {Hovatta}, {Nieppola},
  {Tornikoski}, {L{\"a}hteenm{\"a}ki}, \& {Valtaoja}}]{Ramakrishnan15}
{Ramakrishnan}, V., {Hovatta}, T., {Nieppola}, E., {et~al.} 2015, \mnras, 452,
  1280

\bibitem[{{Ramakrishnan} {et~al.}(2016){Ramakrishnan}, {Hovatta}, {Tornikoski},
  {Nilsson}, {Lindfors}, {Balokovi{\'c}}, {L{\"a}hteenm{\"a}ki}, {Reinthal}, \&
  {Takalo}}]{Ramakrishnan16}
{Ramakrishnan}, V., {Hovatta}, T., {Tornikoski}, M., {et~al.} 2016, \mnras,
  456, 171

\bibitem[{{Rani} {et~al.}(2017){Rani}, {Krichbaum}, {Lee}, {Sokolovsky},
  {Kang}, {Byun}, {Mosunova}, \& {Zensus}}]{Rani17}
{Rani}, B., {Krichbaum}, T.~P., {Lee}, S.~S., {et~al.} 2017, \mnras, 464, 418

\bibitem[{{Rani} {et~al.}(2014){Rani}, {Krichbaum}, {Marscher}, {Jorstad},
  {Hodgson}, {Fuhrmann}, \& {Zensus}}]{Rani14}
{Rani}, B., {Krichbaum}, T.~P., {Marscher}, A.~P., {et~al.} 2014, \aap, 571, L2

\bibitem[{{Rasmussen} \& {Williams}(2006)}]{Rasmussen06}
{Rasmussen}, C.~E. \& {Williams}, C. K.~I. 2006, {Gaussian Processes for
  Machine Learning}

\bibitem[{{Richards} {et~al.}(2011){Richards}, {Max-Moerbeck}, {Pavlidou},
  {King}, {Pearson}, {Readhead}, {Reeves}, {Shepherd}, {Stevenson},
  {Weintraub}, {Fuhrmann}, {Angelakis}, {Zensus}, {Healey}, {Romani}, {Shaw},
  {Grainge}, {Birkinshaw}, {Lancaster}, {Worrall}, {Taylor}, {Cotter}, \&
  {Bustos}}]{Richards11}
{Richards}, J.~L., {Max-Moerbeck}, W., {Pavlidou}, V., {et~al.} 2011, \apjs,
  194, 29

\bibitem[{{Robertson} {et~al.}(2015){Robertson}, {Gallo}, {Zoghbi}, \&
  {Fabian}}]{Robertson15}
{Robertson}, D.~R.~S., {Gallo}, L.~C., {Zoghbi}, A., \& {Fabian}, A.~C. 2015,
  \mnras, 453, 3455

\bibitem[{{Ros} \& {Lobanov}(2001)}]{Ros01}
{Ros}, E. \& {Lobanov}, A.~P. 2001, in 15th Workshop Meeting on European VLBI
  for Geodesy and Astrometry, ed. D.~{Behrend} \& A.~{Rius}, Vol.~15, 208

\bibitem[{{Savolainen} {et~al.}(2002){Savolainen}, {Wiik}, {Valtaoja},
  {Jorstad}, \& {Marscher}}]{Savolainen02}
{Savolainen}, T., {Wiik}, K., {Valtaoja}, E., {Jorstad}, S.~G., \& {Marscher},
  A.~P. 2002, \aap, 394, 851

\bibitem[{{Scharw{\"a}chter} {et~al.}(2013){Scharw{\"a}chter}, {McGregor},
  {Dopita}, \& {Beck}}]{Scharwaechter13}
{Scharw{\"a}chter}, J., {McGregor}, P.~J., {Dopita}, M.~A., \& {Beck}, T.~L.
  2013, \mnras, 429, 2315

\bibitem[{{Sironi} {et~al.}(2013){Sironi}, {Spitkovsky}, \& {Arons}}]{Sironi13}
{Sironi}, L., {Spitkovsky}, A., \& {Arons}, J. 2013, \apj, 771, 54

\bibitem[{{Strauss} {et~al.}(1992){Strauss}, {Huchra}, {Davis}, {Yahil},
  {Fisher}, \& {Tonry}}]{Strauss92}
{Strauss}, M.~A., {Huchra}, J.~P., {Davis}, M., {et~al.} 1992, \apjs, 83, 29

\bibitem[{{Valtaoja} {et~al.}(1992){Valtaoja}, {Terasranta}, {Urpo},
  {Nesterov}, {Lainela}, \& {Valtonen}}]{Valtaoja92}
{Valtaoja}, E., {Terasranta}, H., {Urpo}, S., {et~al.} 1992, \aap, 254, 71

\bibitem[{{van der Laan}(1966)}]{vdLaan66}
{van der Laan}, H. 1966, \nat, 211, 1131

\bibitem[{{Vaughan}(2005)}]{Vaughan05}
{Vaughan}, S. 2005, \aap, 431, 391

\bibitem[{Virtanen {et~al.}(2020)Virtanen, Gommers, Oliphant, Haberland, Reddy,
  Cournapeau, Burovski, Peterson, Weckesser, Bright, {van der Walt}, Brett,
  Wilson, Millman, Mayorov, Nelson, Jones, Kern, Larson, Carey, Polat, Feng,
  Moore, {VanderPlas}, Laxalde, Perktold, Cimrman, Henriksen, Quintero, Harris,
  Archibald, Ribeiro, Pedregosa, {van Mulbregt}, \& {SciPy 1.0
  Contributors}}]{2020SciPy-NMeth}
Virtanen, P., Gommers, R., Oliphant, T.~E., {et~al.} 2020, Nature Methods, 17,
  261

\bibitem[{{Wajima} {et~al.}(2020){Wajima}, {Kino}, \& {Kawakatu}}]{Wajima20}
{Wajima}, K., {Kino}, M., \& {Kawakatu}, N. 2020, \apj, 895, 35

\bibitem[{{Walker} {et~al.}(2000){Walker}, {Dhawan}, {Romney}, {Kellermann}, \&
  {Vermeulen}}]{Walker00}
{Walker}, R.~C., {Dhawan}, V., {Romney}, J.~D., {Kellermann}, K.~I., \&
  {Vermeulen}, R.~C. 2000, \apj, 530, 233

\bibitem[{{Witzel} {et~al.}(2021){Witzel}, {Martinez}, {Willner}, {Becklin},
  {Boyce}, {Do}, {Eckart}, {Fazio}, {Ghez}, {Gurwell}, {Haggard},
  {Herrero-Illana}, {Hora}, {Li}, {Liu}, {Marchili}, {Morris}, {Smith},
  {Subroweit}, \& {Zensus}}]{Witzel21}
{Witzel}, G., {Martinez}, G., {Willner}, S.~P., {et~al.} 2021, \apj, 917, 73

\end{thebibliography}

\begin{appendix} 

\section{GPR analysis} \label{App:GPR}

\subsection{Flares for time lag calculation}\label{App:GPR_flare}

In Fig.~\ref{fig:GP}, the dark-green arrows show the (major) flare used to calculate the time lag between the same as in Fig.~\ref{fig:DCF} light curve frequency pairs.
For the 15.0 (OVRO), 37.0, 91.5, and 230\,GHz light curves, a secondary (minor) flare could robustly be detected (and is denoted in Fig.~\ref{fig:GP} with a dark-red arrow).
We therefore used the average time lag between the major and minor flares to compute the time lag between the cross-correlation pairs 37.0-15.0\,GHz, 91.5-37.0\,GHz, and 230-91.5\,GHz.

\begin{table*}[th]
\begin{center}
\begin{threeparttable}
\caption{Subtracted flux ($S_0$) and PSD power law index ($s$).}             
\label{table:Synth}      
\centering    
\begin{tabular}{ccc}
Observation & $S_0$ & s\\
\hline\hline
 4.8 [GHz]           &  14.35\ [Jy]                                &  2.10\,$\pm$\,0.14\\
 8.0 [GHz]           &  12.99\ [Jy]                                &  1.72\,$\pm$\,0.15\\
 15.0 (UMRAO) [GHz]  &  13.93\ [Jy]                                &  1.35\,$\pm$\,0.13\\
 15.0 (OVRO) [GHz]   &  13.78\ [Jy]                                &  0.02\,$\pm$\,0.09\\
 37.0 [GHz]          &  7.13\ [Jy]                                 &  0.72\,$\pm$\,0.09\\
 91.5 [GHz]          &  14.19\ [Jy]                                &  2.01\,$\pm$\,0.24\\
 230.0 [GHz]         &  1.86\ [Jy]                                 &  0.72\,$\pm$\,0.08\\
 345.0 [GHz]         &  1.33\ [Jy]                                 &  2.57\,$\pm$\,0.13\\
 $\gamma$            &  0.53\ $[10^7\,\times\, \#/cm^{-2} s^{-1}]$ &  1.16\,$\pm$\,0.17\\
\hline
\end{tabular}
\end{threeparttable}
\end{center}
\end{table*}

\section{DCF analysis} \label{App:DCF}

\subsection{Parameters and uncertainties}\label{App:DCF_Parameters}

Applying the DCF implementation on our data sets required two main parameters: the bin size used to search for correlated features and the step size.
For the bin size a conservative approach is to use half the observation duration \citep[see for example][]{Rani17}.
However, in our case, the light curves are not of the same length, and thus we used an asymmetric bin size, which is defined as $[-\max(LC_\textrm{i}, LC_\textrm{j}),\ \min(LC_\textrm{i}, LC_\textrm{j})]$ if the light curve $LC_\textrm{i}$ observations start before the ones of light curve $LC_\textrm{j}$, or  $[-\min(LC_\textrm{i}, LC_\textrm{j}),\ \max(LC_\textrm{i}, LC_\textrm{j})]$, if the opposite is the case.
For the step size $t_\textrm{step}^\textrm{ij}$, we used the following formula:
\begin{equation}
    t_\textrm{step}^\textrm{ij} = \frac{\min(LC_\textrm{i}) + \max(LC_\textrm{j})}{2\mathcal{N}}, \label{eq:tstep}
\end{equation}
where $\mathcal{N}$ is a number given by the following relation:
\begin{equation}
    \mathcal{N} = \min\left(\frac{\mathcal{N}_{LC_\textrm{i}}}{\Delta t_\textrm{i}},\ \frac{\mathcal{N}_{LC_\textrm{j}}}{\Delta t_\textrm{j}} \right), \label{eq:N}
\end{equation} 
with $\mathcal{N}_{LC_\textrm{i/j}}$ being the number of data points and $\Delta t_\textrm{i/j}$ the time range of the observations of $LC_\textrm{i/j}$.
The given uncertainty in each DCF data point in Figs.~\ref{fig:DCF} and \ref{fig:DCF_gamma} was conservatively set to the size of the step size.
A summary of the used parameters is presented in Table~\ref{table:DCF}.

In order to robustly determine the position of the DCF peak, we performed a least squares minimisation Gaussian function fit to the main DCF peak, as shown with the dashed red line in Figs.~\ref{fig:DCF} and \ref{fig:DCF_gamma}.
We note that, for the DCF of the centimetre and millimetre radio light curves, a single Gaussian function sufficiently fitted the data, whereas for the DCF of the 1\,mm and 0.8\,mm with the $\gamma$-ray light curves, two Gaussian functions were required (see also the discussion in Sect.~\ref{sec:Discussion}).
The uncertainty of the fit was then used as an estimate of the time lag uncertainty.
Standard error propagation was used to determine the cumulative error of the time lag between each frequency and the reference frequency (345\,GHz).

\subsection{Confidence interval}\label{ssec:DFC_Confidence}

The DCF peaks in Fig.~\ref{fig:DCF} are all above the 0.95 level, indicating that the correlation is significant.
The same however cannot be said for the DCF peaks in Fig.~\ref{fig:DCF_gamma}.
We therefore calculated the 99.7\% confidence bands, denoted with the dashed golden line in Figs.~\ref{fig:DCF} and \ref{fig:DCF_gamma}.
Specifically, we used the publicly available code presented in \cite{Witzel21}, to create 5000 synthetic light curves \citep{MaxMoerbeck14} based on Gaussian processes, which retain the characteristics of the observed light curves.
The parameters required to create them are the step size, for which we used Eq.~\ref{eq:tstep}, and the power spectral density (PSD).
For the used data sets, a power law model of the form $P\left(f\right) \propto f^{-s}$, with $s$ (see Table~\ref{table:Synth}) being the slope, described the PSD sufficiently well and was thus fitted using a least squares minimisation \citep[see][]{Hodgson18}.
Further details regarding the PSD analysis can be found, for example, in \cite{Vaughan05}, \cite{Rani14}, \cite{Chidiac16}, and \cite{Witzel21}.

\subsection{Light curve DCF time range}\label{ssec:DFC_LC}

As displayed in Fig.~\ref{fig:LCs}, the light curves of \C\ at 4.8, 8.0, 14.8, 37\,GHz span multiple decades, resulting in a very broad DCF curve, if the entire data set is used for the cross-correlation.
If instead the data set is limited around the common flare used as the main cross-correlation feature (denoted with dark-green arrows in Fig.~\ref{fig:GP}), the resulting DCF curve is narrower, with a better defined peak.
We therefore only used the time range of [1978, 1990] for the DCF pairs: 8.0-4.8\,GHz and 14.8-8.0\,GHz, which is centred around the total centimetre-flux peak (see first and second panel, top row in Fig.~\ref{fig:DCF}).
Furthermore, for the DCF pair 37.0-15.0\,GHz we limited the 37\,GHz light curve to the time range [2008, 2022], to match the available time range of OVRO 15\,GHz light curve (see top row, third panel in Fig.~\ref{fig:DCF}).

\begin{table*}[th]
\begin{center}
\begin{threeparttable}
\caption{DCF parameters.}
\label{table:DCF}   
\centering   
\begin{tabular}{ccc}
DCF pair & Bin limits [yrs] & Step size [yrs]\\
\hline\hline
 8.0-4.8 [GHz]         &  [-6, 6]  &  0.1\\
 15.0(UMRAO)-8.0 [GHz] &  [-6, 6]  &  0.1\\
 37.0-15.0(OVRO) [GHz] &  [-6, 7]  &  0.1\\
 91.5-37.0 [GHz]       &  [-7, 3]  &  0.1\\
 230-91.5 [GHz]        &  [-3, 10] &  0.2\\
 345-230 [GHz]         &  [-9, 9]  &  0.7\\
 $\gamma$-230 [GHz]    &  [-7, 10] &  0.1\\
 $\gamma$-345 [GHz]    &  [-7, 9]  &  0.6\\
\hline
\end{tabular}
\end{threeparttable}
\end{center}
\end{table*}

\end{appendix}

\end{document}